\documentclass[sigconf,screen,authorversion]{acmart}

\usepackage{soul}
\colorlet{usercolorname}{yellow!0}
\sethlcolor{usercolorname}

\usepackage{multirow} 
\usepackage{makecell}
\usepackage{CJK}

\usepackage{color}

\AtBeginDocument{%
  }

\copyrightyear{2025}
\acmYear{2025}
\setcopyright{cc}
\setcctype{by-nc-sa}
\acmConference[CHI '25]{CHI Conference on Human Factors in Computing Systems}{April 26-May 1, 2025}{Yokohama, Japan}
\acmBooktitle{CHI Conference on Human Factors in Computing Systems (CHI '25), April 26-May 1, 2025, Yokohama, Japan}\acmDOI{10.1145/3706598.3713709}
\acmISBN{979-8-4007-1394-1/25/04}

\begin{document}
\begin{CJK}{UTF8}{ipxm}

\title[Super Kawaii Vocalics]{Super Kawaii Vocalics: \protect\\ Amplifying the ``Cute'' Factor in Computer Voice}


\author{Yuto Mandai}
\email{mandai.y.aa@m.titech.ac.jp}
\orcid{0009-0002-5866-155X}
\affiliation{
  \institution{Institute of Science Tokyo}
  \city{Tokyo}
  \country{Japan}
}

\author{Katie Seaborn}
\email{seaborn.k.aa@m.titech.ac.jp}
\orcid{0000-0002-7812-9096}
\affiliation{
  \institution{Institute of Science Tokyo}
  \city{Tokyo}
  \country{Japan}
}

\author{Tomoyasu Nakano}
\email{t.nakano@aist.go.jp}
\orcid{0000-0001-8014-2209}
\affiliation{%
  \institution{AIST}
  \state{Ibaraki}
  \country{Japan}
}

\author{Xin Sun}
\email{xin.sun@uva.nl}
\orcid{0000-0002-8188-7576}
\affiliation{%
  \institution{University of Amsterdam}
  \city{Amsterdam}
  \country{Netherlands}
}

\author{Yijia Wang}
\email{wang.y.cf@m.titech.ac.jp}
\orcid{0009-0004-2250-9163}
\affiliation{%
  \institution{Institute of Science Tokyo}
  \city{Tokyo}
  \country{Japan}
}

\author{Jun Kato}
\email{jun.kato@aist.go.jp}
\orcid{0000-0003-4832-8024}
\affiliation{%
  \institution{AIST}
  \state{Ibaraki}
  \country{Japan}
}

\renewcommand{\shortauthors}{Mandai et al.}

\begin{abstract}
  ``Kawaii'' is the Japanese concept of cute, which carries sociocultural connotations related to social identities and emotional responses. Yet, virtually all work to date has focused on the visual side of kawaii, including in studies of computer agents and social robots. In pursuit of formalizing the new science of kawaii vocalics, we explored what elements of voice relate to kawaii and how they might be manipulated, manually and automatically. We conducted a four-phase study (grand \hl{$N=512$}) 
  with two varieties of computer voices: text-to-speech (TTS) and game character voices. We found kawaii ``sweet spots'' through manipulation of fundamental and formant frequencies, but only for certain voices and to a certain extent. Findings also suggest 
  a ceiling effect for the kawaii vocalics of certain voices. We offer empirical validation of the preliminary kawaii vocalics model and an elementary method for manipulating kawaii perceptions of computer voice.
  
\end{abstract}





\maketitle

\section{Introduction}
Advances in human-computer interaction (HCI) are not only contributing to the usability of technology, but also to the user experience (UX). One topic that has recently surged in importance is voice UX~\cite{seaborn_measuring_2021,clark_state_2019,seaborn2021voiceagent,seaborn2024qualvoice}. This field aims to understand and improve the fit between people and computers in situations where the interaction is based on speech~\cite{clark_state_2019,seaborn2021voiceagent}. Creators are building interfaces that enable natural dialogue~\cite{lahiri_hybrid_2023,doyle2021pmq}, notably with voice assistants, like Amazon Alexa, and conversational AI, like ChatGPT. 
Voice can be powerful in certain contexts~\cite{seaborn2024qualvoice}, such as speech-controlled hands-free in-car systems used while driving~\cite{wong_voices_2019}, and, in an increasingly globalized world, providing a means to understand the pronunciation and cultural nuances of different languages~\cite{zhou_chat_2023}. Understanding the UX that voice-based systems provide for a diversity of people is now a critical trajectory~\cite{clark_state_2019,seaborn_measuring_2021,seaborn2021voiceagent}.

One culturally-sensitive factor that may contribute to voice UX is the sociocultural phenomenon of ``kawaii.'' Kawaii is the Japanese word for ``cute'' or ``pretty'' and has been studied from social and linguistic angles~\cite{heng_beyond_2014,nittono_power_2012}. Kawaii, while carrying special properties in the Japanese context, is not unique to Japan, being linked to evolutionary biology relevant to all people. \citet{nittono_two-layer_2016} proposed a two-layer model to explain kawaii as a human emotion with Japanese cultural components~\cite{nittono_two-layer_2016}. For instance, people who completed a specific task after viewing an image with high ``kawaii''  performed better than those who did not~\cite{nittono_power_2012}. 
However, almost all research has related to visual kawaii~\cite{yijia_kawaiicomputing}. 
In a first effort, \citet{seaborn_can_2023} extended the visual kawaii model to voice and vocalics, considering the UX of synthetic speech. In two studies~\cite{seaborn_can_2023,seaborn_game_2023}, they found that social identity factors, such as gender and age, as well as the mechanical factors of fluency and humanlikeness, influenced the perception of voice kawaii. Also, voices considered gender ambiguous and gender neutral were perceived as especially kawaii by certain individuals.
Kawaii vocalics is a new field of study, and there is great room for further research on the phonological characteristics linked to kawaii. Given its links to emotion~\cite{nittono_psychophysiological_2017}, trust~\cite{Shiomi_2023}, persuasion~\cite{nittono_power_2012,birlea2023soft}, and other UX factors~\cite{wang2024kawaiicomp,Ohkura_2023kawaiiengineering}, kawaii vocalics may be an important tool in designing voice UX. As yet, no attempt has been made to manipulate the perception of kawaii as potential voice design material~\cite{sutton2019designmaterial} in or outside of computer systems. Our work adds new insights on how user perceptions can be influenced by manipulating the kawaiiness of voice stimuli. 

Factors affecting kawaii-relevant sex/gender\footnote{Given the state of reporting, i.e., operationalizations and terminology, and assumptions made about nature and nurture, including their distinction or lack thereof, it is not always clear whether sex-as-biology, gender-as-identity, or both is meant~\cite{Tannenbaum_2019}. We use sex/gender in these cases.} and age perceptions have been explored in the field of phonetics. 
Fundamental and first-to-third formant frequencies are thought to be cues in the sex/gender perception of children to adolescents~\cite{perry_gender,skuk_voicegender} and age~\cite{mori_voice2014}. In the absence of acoustic features linked to kawaii, fundamental and formant frequencies are clear starting points for manipulating perceptions of kawaiiness  and social perceptions of voice. \citet{Kasuya1968jpformant}, considering Japanese vowels, found that the pitch and formant structure of main five---a, i, u, e, and o---vary with sex/gender and age, notably in relation to the length of the oral cavity and vocal tract. Specifically, the first and second formants relate to the pronunciation of vowels. In addition to Japanese, other languages can map fundamental vowel and formant frequencies~\cite{kent_static_2018, zhenglai_analysis_2003}, suggesting that kawaii-ness or ``cuteness'' may be manipulable in the Japanese context and possibly other sociocultural contexts. However, this has not been explored.

Here, we report on our four-phase project representing the first attempt to study whether and how kawaii vocalics can be manipulated as voice design material~\cite{sutton2019designmaterial}. Our goal was to identify the nature of kawaii vocalics in terms of audio features and elicited perceptions of social factors. We thus had two intertwined research questions (RQs): \textbf{RQ1.} \emph{What voice features lead to perceptions of voices as kawaii?} and \textbf{RQ2.} \emph{What voice features linked to kawaii are also linked to social identity perceptions?} We started with a simple audio feature search and its validation through manual manipulation and then aimed at an automatic approach through speech signal processing. We then attempted to confirm the results on the first set of voices---text-to-speech (TTS) systems---by using an existing datasets---the game character voices originally studied by \citet{seaborn_game_2023}--to assess the degree to which our kawaii vocalics manipulation techniques could be universal across digital (or digitized) voices, i.e., a ``Super Kawaii Vocalics'' technique. We offer the following major contributions:

\begin{itemize}
    \item Empirical demonstration of manual and, with caveats, automatic methods for manipulating a diversity of digital voice stimuli in relation to perceptions of kawaiiness , in both directions, i.e., \textbf{kawaii±})
    \item Empirical validation of the initial results~\cite{seaborn_can_2023,seaborn_game_2023} linking social identity perceptions to kawaii voice, and how these can shift when kawaii-ness is manipulated
    \item Evidence of voice-specific ``sweet spots'' and a ``ceiling effect'' for voices that may already be peak kawaii
\end{itemize}

We offer this work as a first step towards clarifying what factors of voice contribute to kawaii perceptions by identifying the perceptual changes that result from voice manipulations. The novelty is in developing the psycho-perceptual arm of voice UX research and  awareness and study of sociocultural phenomena outside of the English-speaking West~\cite{Henrich2010}. We contribute to the diverse landscape of cultural studies in computing and beyond~\cite{Kato2023SIGCCC,Ohkura_2023kawaiiengineering,himmelsbach2019we}.


\section{Theoretical Background}
\label{sec:theoreticalbg}

We begin with an overview of how kawaii has been approached as a perceptual phenomenon of study, shifting to the novel topic of kawaii vocalics and how it may be manipulated through audio and signal processing techniques.

\subsection{Theoretical Models for Kawaii Perceptions: Visual and Auditory}
Kawaii is a multiform phenomenon in Japanese culture. As an adjective, ``kawaii'' can mean ``cute,'' ``pretty,'' ``sweet,'' ``adorable,'' and other terms of endearment and appeal. However, \citet{nittono_power_2012} realized that kawaii is also an emotion with social and biological origins deserving of scientific study. Over several defining studies, e.g., \cite{nittono_power_2012,nittono_two-layer_2016,nittono_psychophysiological_2017}, Nittono and colleagues proposed a \emph{two-layer model of kawaii}~\cite{nittono_two-layer_2016}. On one level, kawaii is linked to aspects of Japanese culture, such as ``amae,'' behaviour expressing a desire to be loved by others, and ``chizimi shikou,'' the love of small things~\cite{nittono_two-layer_2016}. Yet, from a biological perspective, kawaii is also linked to how the brain processes social information, suggesting its universality. This cognitive phenomenon is called Kindchenschema (baby schema)~\cite{lorenz_angeborenen_1943}. This refers to visual characteristics of young animals, such as a large head and eyes compared to the body, roundness of shape and soft texture, etc., which stimulate a care response. Essentially, baby-like stimuli invoke certain responses in the human brain geared around protection and endearment, likely for the survival of the species, because infants are resource-intensive. \citet{nittono_psychophysiological_2017} showed that this cross-cultural notion of baby schema acts as trigger for the emotion of kawaii in Japanese people. Pertinent to our work, \citet{nittono_creation_2022} began exploring whether kawaii responses can be influenced by manipulating aspects of the visual stimulus. For this the researchers visually mixed infant faces to create an intermediate kawaii infant face that shifted kawaii ratings~\cite{nittono_creation_2022}. We predicted that a similar phenomenon could be produced for voice.

The science of \emph{kawaii vocalics} was created in response to the dearth of work on kawaii and sound modalities~\cite{seaborn_can_2023,seaborn_game_2023}. In the first paper, \citet{seaborn_can_2023} extended the visual-oriented model of kawaii by \citet{nittono_two-layer_2016}, investigating ``voice'' as kawaii sound and directly addressing the link to social identities, given the cultural stereotype of kawaii as ``girlish''~\cite{shiokawa1999cute}. In a user perception study of TTS voice clips, they discovered that voices deemed ``girlish'' but also ``gender ambiguous,'' e.g., having a mixture of feminine and masculine characteristics, and anthropomorphic (via fluency and humanlikeness) were also the most kawaii. This work was purely descriptive and focused on TTS voices.
In their follow-up work, \citet{seaborn_game_2023} applied the preliminary model of kawaii vocalics to a new sample: 18 popular video game characters in Japan. Notably, the quality of the samples was much higher, given that TTSs are generative and limited by the training and models, and video game character voices are performed by professional voice actors and/or refined through professional processing. The results confirmed the original model and extended its reach to a new gender factor: ``gender neutrality,'' where ``genderless'' (neither masculine nor feminine) voices were also deemed most kawaii. The notable example was Pikachu from the Pok\'{e}mon series. However, as before, this work was purely descriptive and theory-generating, with no manipulation of kawaii vocalics aside from stimuli selection.

We used these two foundational studies of kawaii vocalics~\cite{seaborn_can_2023,seaborn_game_2023} as a theoretical and methodological baseline. We aimed to extend this work by exploring how kawaii perceptions can be manipulated by modifying (and then evaluating) perceptions of the key factors that emerged from both studies: voice gender, age, and anthropomorphism. We used the same voice samples for comparison of the original, unaltered voices and the manipulated voices in our study.
For clarity in reporting, we define these key terms:
\begin{itemize}
    \item \textbf{Kawaii}: The socioemotional phenomena~\cite{nittono_power_2012}.
    \item \textbf{Kawaiiness}: The property of a stimulus that elicits perceptions and feelings of kawaii that may be manipulated.
    \item \textbf{Kawaii perceptions}, \textbf{kawaii ratings}, \textbf{perceptions of kawaiiness}, or \textbf{perceived kawaiiness}: The socioemotional response from an individual to the stimulus.
    \item \textbf{Kawaii vocalics}: Features of voice stimuli that convey meaning, including social identity cues, and emotion~\cite{seaborn_can_2023}.
    \item \textbf{Kawaii voice}: Voice stimuli perceived as kawaii.
\end{itemize}

\subsection{Manipulating Vocalics through Manual and Automated Speech Signal Processing}
Research on speech features and feature recognition has been conducted across a range of fields, from signal processing to biology to acoustics to HCI. \citet{jain2022protosound}, for instance, explored how to allow deaf and hard of hearing people to create their own custom models of sound phenomena for machine learning-based sound recognition, a form of audio manipulation. Kawaii voice perceptions are based on specific sociocultural models and individual experiences of specific voices, including sociolinguistic and paralanguage features like pronunciation~\cite{shiokawa1999cute, seaborn_can_2023, Nittono2021crosscultural, lieber2021otona, yomota2006_kawaiiron}. In the kawaii vocalics model, girlishness, gender ambiguity, and young age relate to kawaii~\cite{seaborn_can_2023}. These social perceptions are tied to the fundamental ($F0$) and formant frequencies of voice. $F0$ is a physical quantity related to the pitch of a sound. $F0$ decreases with secondary sexual characteristics in humans~\cite{childers_gender}, with males ultimately having lower $F0$ than females~\cite{huber_formant,childers_gender}. 
Fundamental frequency is an important factor in discriminating between male and female voices~\cite{pisanski2011_prior}.
Formant frequency also varies with the length of the vocal tract~\cite{kent2002acoustic}.
The vocal tract modulates the resonant frequency, or formant~\cite{tanaka2022effects}.
The male vocal tract tends to be longer than the female vocal tract~\cite{huber_formant}, so formant frequency is likewise an important factor in discriminating between male and female voices. Given its links to gender and age perceptions~\cite{seaborn_can_2023,seaborn_game_2023,perry_gender, skuk_voicegender,mori_voice2014}, we started our exploration with
$F1$--$F3$ formant frequencies. 
We hypothesized:

\begin{quote}
    \textbf{H1:} Higher fundamental and formant frequencies will increase perceptions of kawaiiness .
\end{quote}

\citet{seaborn_can_2023} concluded that kawaii vocalics is related to perceptions of a young age. Therefore, manipulating voice youthfulness may lead to an increase in perceived kawaiiness. \citet{Kasuya1968jpformant} showed that $F0$ descended with increasing age, particularly in female speech age perceptions. \citet{huber_formant} also showed that the first, second, and third formant frequencies changed in relation to age and gender, roughly in the opposite direction of age and frequency. In kind, we hypothesized:

\begin{quote}
    \textbf{H2a:} Higher fundamental and formant frequencies will result in younger age perceptions.
\end{quote}

Multidisciplinary work differentiated sex (biological characteristics) and gender (social identity)~\cite{Tannenbaum_2019}, as well as raised awareness of transgender and intersex people~\cite{Dunham_2019transinter,Spiel_2019patchinggender}. Most work is premised in the gender binary model of male/female or masculine/feminine, which is often applied to humanlike but non-human subjects like machines~\cite{Perugia_2023gender}. \citet{seaborn_can_2023,seaborn_game_2023}, taking a ``gender-expansive'' approach for computer agents~\cite{seaborn2022expansive}, found novel results for gender: the TTS and game character voices perceived as girlish, gender ambiguous, or gender-neutral were considered most kawaii. A notable example was Pikachu and other non-human (but non-mechanical) characters. This led us to hypothesize:

\begin{quote}
    \textbf{H2b:} Higher fundamental and formant frequencies will result in ambiguous or gender-neutral perceptions.
\end{quote}

\emph{Automation} is a key step in \emph{applying} fundamental user perceptions and manual manipulation work. Once 
understood, manipulation of kawaii vocalics can be automated through computational signal processing.
Our goal was to contribute to the \emph{practice} of kawaii vocalics. While confirming previous work is important, we also wished to show how the findings can be \emph{used} by researchers, professionals, and potentially laypeople. By changing the fundamental and formant frequencies through speech signal processing, it is possible to manipulate the kawaii \emph{and} social perceptions of a given voice at the same time, in a controlled way. This led us to develop and test automated procedures (refer to \autoref{sec:phase1}).

A variety of methods exist for manipulation in speech signal processing. In the field of music production, a graphical user interface (GUI)-based software called DAW (Digital Audio Workstation) is often used for manual manipulation~\cite{marrington2017composing}. The drawback is it requires manual operations by a sound engineer. A code-based approach 
that could be automated would be ideal for generalizing the method, removing the need for human labour. 
Legacy-STRAIGHT~\cite{kawahara1997straight, KAWAHARA1999187} is a widely used vocoder. 
WORLD~\cite{MasanoriMORISE20162015EDP7457_WORLD, kawahara1997straight} is also a vocoder often used for speech synthesis and voice conversion.
Both methods have been used extensively in speech research and are popular code-based methods for automated speech signal processing. 
For breadth and flexibility in future applications, we explore both tools here.

\section{Overview of Phases}


We used a multi-phase approach to explore the manipulation of kawaii vocalics. The phases were:

\begin{itemize}
    \item \textbf{Phase 1: Manual Processing with DAW (Cubase) / Audio Signal Processing and TTS Voices (\autoref{sec:phase1})} \\ We first conducted a user perceptions study ($n=50$) of the manually processing approaches expected to be related to kawaii vocalics using TTS-based computer voices ($n=5$). Second, a supplementary study ($n=261$) testing the fine-grained parameters involved in automatic speech signal processing was conducted in pursuit of replicability, convenience, and flexibility.
    We also intended to confirm the theoretical relationship between kawaii, girlishness, and gender/age ambiguity (refer to \autoref{sec:theoreticalbg}) through concrete audio manipulations, and thus set a baseline for more advanced processing and subsequent evaluations.\\
    
    \item \textbf{Phase 2: Application of Speech Signal Processing to Game Character Voices (\autoref{sec:phase3})} \\
    We then applied the manipulation to a larger number of different voices. We used the diverse set of game character voices ($n=18$) from \citet{seaborn_game_2023}.
    We aimed to verify the kawaii vocalics manipulation for a range of voices, evaluated in a user perceptions study ($n=150$).\\

    \item \textbf{Phase 3: Fine-Grained Manipulation of Game Character Voices (\autoref{sec:phase4})} \\ Phase 2 revealed the complexity of automated kawaii voice manipulations. We thus returned to the manual kawaii vocalics manipulation procedure used in Phase 1, as it allowed for customizations per voice and higher quality output. We investigated user perceptions in an online study ($n=51$).\\
\end{itemize}

All phases received ethics approval from the university ethics board (\#2023327). Our protocol was registered on OSF\footnote{\url{https://osf.io/9eq7w}} before data collection on July 15\textsuperscript{th}, 2024.

\section{Phase 1: Manual Processing with DAW (Cubase) / Audio Signal Processing and TTS Voices}
\label{sec:phase1}

Fundamental and formant frequencies are factors in estimating the sociodemographic attributes of the person speaking~\cite{perry_gender,mori_voice2014}. We can expect that a high-pitched voice will be perceived as being young and girlish, which is a factor in the kawaii vocalics model~\cite{seaborn_can_2023}.
Therefore, using TTS voices, we investigated how a simple fundamental/formant frequency ``rising'' manipulation could affect kawaii perceptions. 
First, we aimed to show that voices with higher fundamental or formant frequencies produced by the manipulation would actually be perceived as high-pitched voices. Second, we aimed to reveal the relationship between perceptions of voice height and kawaii.
We conducted an exploratory study of the manual voice manipulations with DAW (labeled 1) alongside a study to ensure the usefulness of automatizing these manipulations (labeled 2). Notably, we compared the user perceptions data by manual and automated manipulation.

\subsection{Participants}
\begin{enumerate}
    \item Participants ($N=50$, women $n=11$, men $n=39$, another gender or N/A $n=0$) were recruited through Yahoo! Crowdsourcing Japan on December 15\textsuperscript{th}, 2023. This participant pool platform, similar to Prolific or Amazon Mechanical Turk, is among the largest in Japan, given the reach of Yahoo! within the country. Accounts are linked to Yahoo! JAPAN IDs and specific phone numbers, with reregistration only possible after six months\footnote{\url{https://crowdsourcing.yahoo.co.jp/agent/guideline} (Note: Japanese only)}. This ensures certain basic demographics and reduces the likelihood of multiple accounts. Most respondents were between 35-44 ($n=19$) and 45-54 ($n=14$) years of age, with some younger (18-34, $n=10$) and older (55-74, $n=7$). Most did not use a VA ($n=25$), but many used one daily or weekly ($n=18$). Five participants used a VA once a month and two used a VA previously but did not anymore. Participants were paid $\sim$600 yen for 30 minutes, according to ethics guidelines.
    
    \item Participants ($N=50$, women $n=6$, men $n=43$, another gender or N/A $n=1$) through Yahoo! Crowdsourcing Japan on July 8, 2024. Most respondents were between 35-44 ($n=20$) and 45-54 ($n=16$) years of age, with some younger (25-34, $n=2$) and older (55-74, $n=9$). Participants were compensated $\sim$600 yen for 30 minutes as per ethics guidelines.
\end{enumerate}

\subsection{Procedure}
Participants were given a link to a SurveyMonkey\footnote{\url{https://www.surveymonkey.com}} questionnaire on the Yahoo! recruitment page. After giving consent, they were presented with a joint audio and attention check, where they had to input a number spoken aloud by a neutral voice to proceed. They were then presented with 5-second voice clips---25 clips in (1) and 15 clips in (2)---generated with voice assistant (VA) TTSs (refer to \ref{sec:p1materials}). For each clip, they provided their impressions through self-report prompts (refer to \ref{sec:p1measures}).
To counter novelty and order effects, the clips were presented in random order~\cite{schuman1996questions}. Participants entered demographics on the last page. Both studies took $\sim$30 minutes.

\subsection{Materials}
\label{sec:p1materials}

\subsubsection{Voice Stimuli}
\begin{figure*}[!ht]
    \centering
    \includegraphics[width=.6\textwidth]{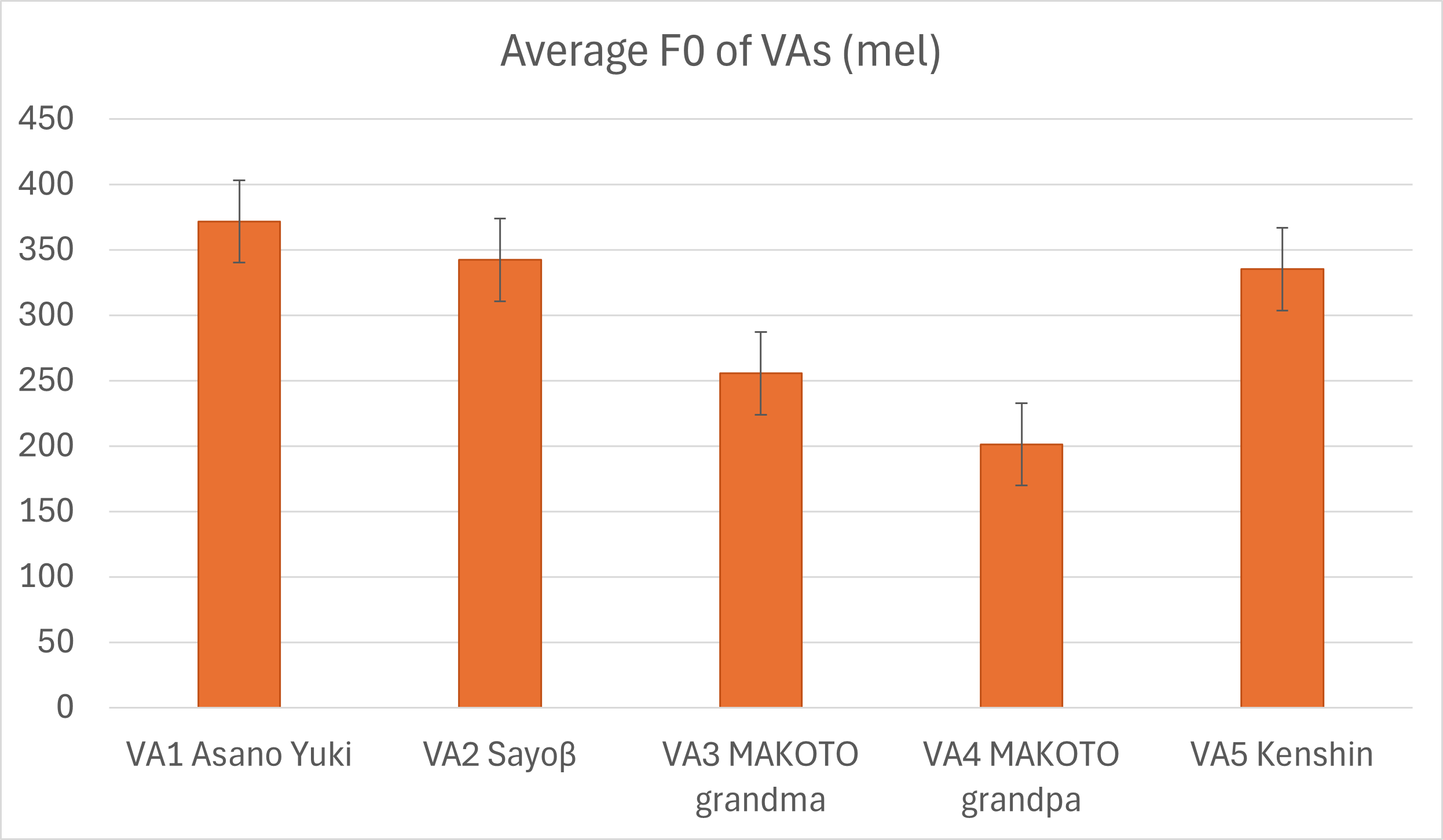}
    \caption{Mean fundamental frequencies across the five TTS voices  (Phase 1).}
    \label{fig:p1mean}
    \Description{Figure of the mean fundamental frequencies for each TTS voices, showing that each voice has different pitched voices, with young woman Asano Yuki having the highest and grandpa MAKOTO having the lowest.}
 \end{figure*}

We used five voices. Three---Sayo$\beta$, Asano Yuki, and Kenshin---were from CoeFont\footnote{\url{https://coefont.cloud}}, a Japanese TTS provider. Two were novel ``older adult'' TTSs, a set of ``grandma'' and ``grandpa'' voices that have not been publicly released. The voices were generated from a novel TTS created with $\sim$5-7 hours of recordings from two older adults (aged 65+).
Each voice had a different average pitch; the details are presented in \autoref{fig:p1mean}.

\begin{enumerate}
    \item For each voice, five patterns of fundamental and formant frequencies were created (\autoref{tab:p1voicemanip}) in Steinberg Cubase Pro 9.5\footnote{\url{https://www.steinberg.net/cubase}}. 
    The fundamental and formant frequencies were set to $2^{\frac{3}{12}}$ for higher frequencies and $2^{-\frac{3}{12}}$ for lower frequencies. This corresponds to the twelve mean rule of music theory based on piano and keyboard instruments \cite{suga_utilize}. As per \citet{barbour2004tuning}, in music theory, one octave is divided into twelve semitones (12 equal temperaments). If the sound goes up one semitone, then the frequency becomes $2^{\frac{3}{12}}$ times higher. 
    In this study, each voice was moved up or down by three semitones. 
    This resulted in a total of 25 voice clips. The clips were uploaded to SoundCloud\footnote{\url{https://soundcloud.com/discove}}, a free streaming platform that allows embedding of sound clips in online questionnaires. 

    \item The same TTS voices as in (1) were used.
    Voice clips with fundamental and spectrogram frequencies three semitones higher were prepared for each of the five voices using three voice synthesis methods. One was the manual approach (via Cubase by Steinberg) used in (1). The others were the code-based Legacy-STRAIGHT by \citet{kawahara1997straight} and WORLD by \citet{MasanoriMORISE20162015EDP7457_WORLD}.
    Legacy-STRAIGHT and WORLD decompose the input speech signal into the fundamental frequency, the spectral envelope, and the aperiodicity as a set of vocoder parameters~\cite{MasanoriMORISE20162015EDP7457_WORLD}.
    Pitch height can be changed through the fundamental frequency and the frequency response (\textit{i.e.}, voice timbre), including formant frequencies, by manipulating the spectral envelope.
\end{enumerate}

\begin{table*}[!ht]
\caption{Overview of VA voice clips and manually applied manipulations in Cubase (Phase 1).}
\label{tab:p1voicemanip}
\begin{tabular}{@{}llrrrr@{}}
\toprule
   Clip ID &Voice & Fundamental Freq. & 1st$\sim$3rd Formant Freq. \\
\midrule
   VA101 & あさのゆき (Asano Yuki; CoeFont) & ±0 (semitone) & ±0 (semitone) \\
   VA102 & あさのゆき (Asano Yuki; CoeFont) & +3 & +3 \\
   VA103 & あさのゆき (Asano Yuki; CoeFont) & -3 & -3 \\
   VA104 & あさのゆき (Asano Yuki; CoeFont) & ±0 & +3 \\
   VA105 & あさのゆき (Asano Yuki; CoeFont) & ±0 & -3 \\
   VA201 & 小夜$\beta$ (Sayo$\beta$; CoeFont) & ±0 & ±0 \\
   VA202 & 小夜$\beta$ (Sayo$\beta$; CoeFont) & +3 & +3 \\
   VA203 & 小夜$\beta$ (Sayo$\beta$; CoeFont) & -3 & -3 \\
   VA204 & 小夜$\beta$ (Sayo$\beta$; CoeFont) & ±0 & +3 \\
   VA205 & 小夜$\beta$ (Sayo$\beta$; CoeFont) & ±0 & -3 \\
   VA301 & おばあさん (grandmother; MAKOTO) & ±0& ±0 \\
   VA302 & おばあさん (grandmother; MAKOTO) & +3 & +3 \\
   VA303 & おばあさん (grandmother; MAKOTO) & -3 & -3 \\
   VA304 & おばあさん (grandmother; MAKOTO) & ±0 & +3 \\
   VA305 & おばあさん (grandmother; MAKOTO) & ±0 & -3 \\
   VA401 & おじいさん (grandmother; MAKOTO) & ±0 & ±0 \\
   VA402 & おじいさん (grandmother; MAKOTO) & +3 & +3 \\
   VA403 & おじいさん (grandmother; MAKOTO) & -3 & -3 \\
   VA404 & おじいさん (grandmother; MAKOTO) & ±0 & +3 \\
   VA405 & おじいさん (grandmother; MAKOTO) & ±0 & -3 \\
   VA501 & けんしん (Kenshin CoeFont) & ±0 & ±0 \\
   VA502 & けんしん (Kenshin CoeFont) & +3 & +3 \\
   VA503 & けんしん (Kenshin CoeFont) & -3 & -3 \\
   VA504 & けんしん (Kenshin CoeFont) & ±0 & +3 \\
   VA505 & けんしん (Kenshin CoeFont) & ±0 & -3 \\
   \bottomrule
 \end{tabular}
\end{table*}

\subsubsection{Speech Content}
We used the Japanese translations from \citet{seaborn_can_2023} of the English phrases in \citet{baird_perception_2017}:「おげんきですか。ありがとうございます。あなたを愛しています。」 (``How are you? Thank you. I love you.'')
We used the same sentences for potential future comparisons. 

\subsection{Measures and Instruments}
\label{sec:p1measures}
We used SurveyMonkey, a professional online questionnaire platform. Unless noted, all items were presented in a matrix with responses gathered using 7-point Likert scales: 1: 全くそう思わない (strongly disagree),
2: あまりそう思わない (disagree),
3: どちらかというとそう思わない (somewhat disagree),
4: どちらでもない (n/either),
5: どちらかというとそう思う (somewhat agree),
6: かなりそう思う (agree),
7: 全くそう思う (strongly agree). Item order was randomized to avoid order effects~\cite{schuman1996questions}.

\subsubsection{Kawaii Perceptions}
The one-item scale from \citet{seaborn_can_2023,seaborn_game_2023} was used to determine how ``kawaii'' the respondents felt each voice clip was. We used this measure to ensure comparison with the original datasets, but also in the absence of a validated measure of kawaii for voice.

\subsubsection{Gender Perceptions}
Gender perceptions were examined in two ways. One was captured in a nominal scale consisting of:  feminine (女性的), masculine (男性的), I hear both, i.e., ambiguous (男性的にも女性的にも聞こえる), and I hear neither, i.e., genderless (男性的にも女性的にも聞こえない). Participants were also free to enter alternative options and explanations in an open text field. Individual items were also folded into the Likert scale matrix: feminine (女性的だ), masculine (男性的だ), gender unclear (性別があいまいだ), and no gender (性別をもっていない).

\subsubsection{Age Perceptions}
Agedness was captured, as in the initial work~\cite{seaborn_can_2023,seaborn_game_2023}, through a nominal scale of infant/baby (0-2 years), child (3-12 years), teenaged (13-19 years), adult (20-39 years), middle-aged (40-64 years), older adult (65+ years), and ageless.

\subsubsection{Perceptions of Humanlikeness, Artificiality, and Fluency}
We used the same items as in \citet{seaborn_can_2023, seaborn_game_2023}: 人間らしい (humanlike), 機械的/人工的だ (mechanical/machinelike), and 流暢だ (fluent).
We added 動物的だ (animal-like) and 本物のようだ (authentic) to the automated manipulation study (2) to compare with the planned game characters voice study (\autoref{sec:phase3}). We reasoned that some game characters and species were neither ``human'' nor ``artificial'' or ``machinelike,'' like Pok\'{e}mon, and some characters spoke with odd speech patterns, notably gibberish~\cite{seaborn_game_2023}. The TTS categories were insufficient to describe the nature of such voices.

\subsection{Data Analysis}

\subsubsection{Voice Analysis}
The voice analysis software Praat was used
~\cite{stylerpraat}. Fundamental frequency was assessed using the ``Get Pitch'' command, and formant frequency was analyzed using the ``To Formant (Burg)'' command, which uses the Burg method~\cite{karigome_use_2018}. The maximum, minimum, and average values of the fundamental and time-series frequencies of $F1$--$F3$ 
were assessed. 
Here, considering human perceptual characteristics, the linear frequency $f_\mathrm{Hz}$ is converted to a mel-scale frequency $f_\mathrm{Mel}$ as follows:

\begin{equation} 
f_\mathrm{Mel} = 2595 \log_{10}{\left(1+\frac{f_\mathrm{Hz}}{700}\right)}
\end{equation}

Frequency does not linearly match subjective pitch height and Hz.
Thus, we used the mel-scale to match the auditory characteristics.
For the formants, the average value was calculated except the time when the clip was silent and for the time when the clip was not sampled to $F3$; these were considered to be 
the representative $F1$, $F2$, $F3$. Praat was used to plot the speech waveform, time-series curve of the fundamental frequency, and time-series plot of the formants for each voice (refer to the Supplementary Materials).
These confirmed the success of the fundamental frequency manipulation and the first-to-third formant frequency manipulation applied in Cubase to each of the five voices.

\subsubsection{Statistical Analysis}
\begin{enumerate}
    \item Descriptive statistics---mean ($M$), median ($MD$), standard deviation ($SD$), and interquartile range ($IQR$)---were calculated for the Likert scale data. Shapiro-Wilk tests showed that the data distribution was not normal. As such, Spearman's rank correlation coefficients were generated to relate this data to the voice frequency data. 
    For example, あさのゆき (Asano Yuki) was assigned the label ``VA1.'' Clips VA101, VA102, VA103 were used for analysis of fundamental frequencies, while VA101, VA104, and VA105 were used for analysis of formant frequencies. This division allowed us to isolate effects related to each type of frequency.

    \item Descriptive statistics were created. Shapiro-Wilk tests showed that the data were not normally distributed, so nonparametric Wilcoxon signed-rank tests were used to examine differences between Cubase, STRAIGHT, and WORLD.
\end{enumerate}

\subsection{Results}
\subsubsection{Study (1): Manual Manipulation with DAW}
\label{sec:p1results}

Several voices (\autoref{fig:p1mean}) were perceived as kawaii: the young boy けんしん (Kenshin) VA502-F0$\sim$F3 +3 ($M=4.6, SD=1.9, MD=5, IQR=1$), the young girl
小夜 (SAYO$\beta$) VA201-$\pm0$ ($M=4.4, SD=2.3, MD=5, IQR=1.75$), and also 小夜\ (SAYO$\beta$) VA202-F0$\sim$F3 +3 ($M=4.4, SD=1.4, MD=5, IQR=2$).
Descriptive statistics are listed in the Supplementary Materials.

\begin{figure*}[!ht]
    \centering
    \includegraphics[width=.85\textwidth]{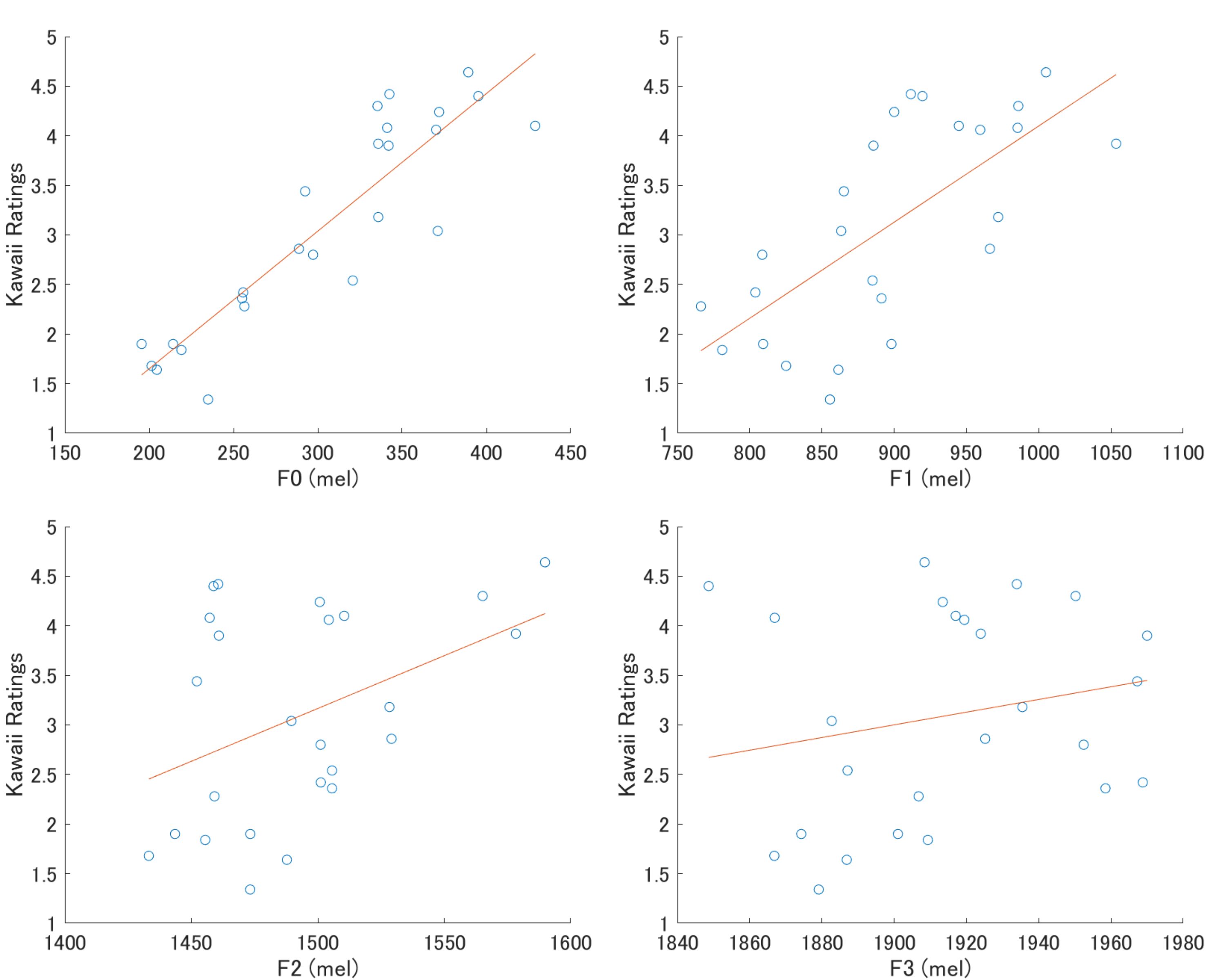}
    \caption{Scatter plot between kawaii perceptions and formants $F0$, $F1$, $F2$, $F3$ (Phase 1).}
    \label{fig:p1scatter_kawaii}
    \Description{Scatter plots between kawaii perceptions and $F0$, $F1$, $F2$, and $F3$. There are strong positive correlations except for $F3$.}
\end{figure*}

A positive, statistically significant Spearman's correlation was found between kawaii perceptions and fundamental frequencies, $r_s(13) =.89, p < .001$, 95\% $ CI  [0.71, 0.96]$. 
Furthermore, both the relationships between kawaii perceptions and first ($r_s(13) = .74, p < .001$, 95\% $ CI  [0.41, 0.89]$) formant frequencies were statistically significant. 
However, no relationship was found between kawaii perceptions and second ($r_s(13) = .35, p = .08$, 95\% $ CI  [-0.076, 0.68]$), third formant frequencies ($r_s(13) = .23, p = .25$, 95\% $ CI [-0.20, 0.60]$).
Therefore, we can partially accept the hypothesis: 

\begin{quote}
    \textbf{H1:} Higher fundamental/formant frequencies increase perception of kawaii.\\ \textbf{$\rightarrow$ Partially accepted (for higher fundamental and first formant frequencies).}
\end{quote}

A negative, statistically significant correlation was found between age perceptions and the fundamental, ($r_s(13) = -.72, p <.001$, 95\% $CI  [-0.87, -0.46]$)  first formant ($r_s(13) = -.85, p < .001$, 95\% $CI  [-0.93, -0.69]$) frequencies. However, none was found for second ($r_s(13) = -.38, p = .06$, 95\% $CI  [-0.67, 0.015]$) and third formant frequencies ($r_s(13) = -.20$, $p = .32$, 95\% $CI[-0.56, 0.21]$).
Therefore, we can partially accept the hypothesis: 

\begin{quote}
    \textbf{H2a:} Higher fundamental/formant frequencies result in younger age perceptions.\\ \textbf{$\rightarrow$ Partially accepted (for higher fundamental and first formant frequencies).}
\end{quote}

Spearman's correlation between gender ambiguity perceptions and fundamental frequencies was not statistically significant. ($r_s(13) = .16, p = .43$, 95\% $CI[-0.25, 0.53]$). 
Similarly, both the relationship between gender ambiguity perceptions and the first ($r_s(13) = .35, p = .08$, 95\% $CI[-0.047, 0.66]$), second  ($r_s(13) = .37, p = .07$, 95\% $CI[-0.027, 0.67]$)formant frequencies were not statistically significant. However, a positive statistically significant correlation was found between gender ambiguity perceptions and the third formant frequencies ($r_s(13) = .47, p = .02$, 95\% $CI[0.10, 0.73]$) 
Therefore, we can partially accept the hypothesis: 
\begin{quote}
    \textbf{H2b:} Higher fundamental/formant frequencies result in ambiguous gender perceptions.\\ \textbf{$\rightarrow$ Partially accepted (only for higher third formant frequencies).}
\end{quote}

\begin{figure*}[!ht]
    \centering
    \includegraphics[width=\textwidth]{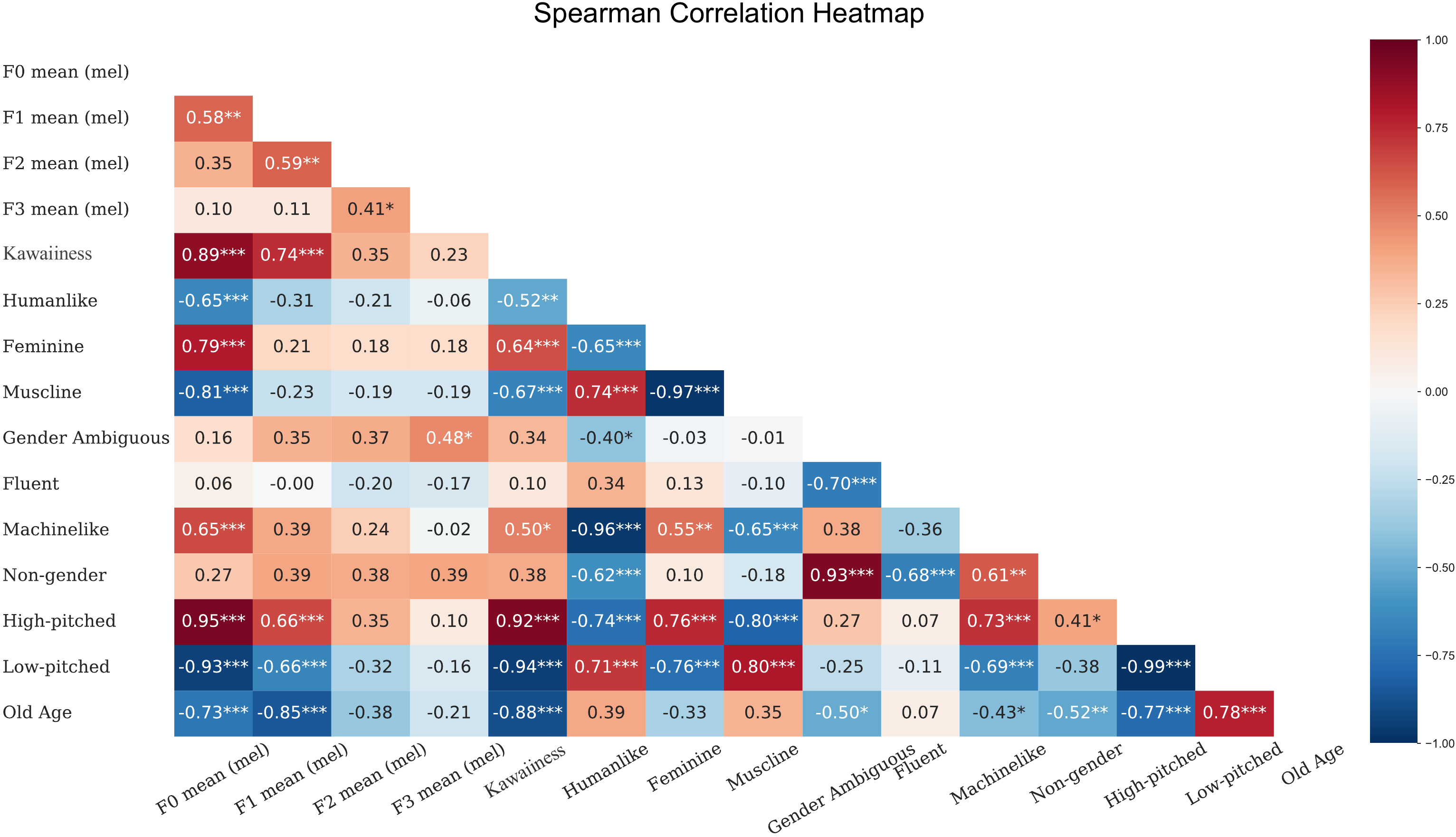}
    \caption{Spearman correlation test results for all variables (Phase 1). $***p<0.001, **p<0.01, *p<0.05$}
    \label{fig:p1correlation}
    \Description{Figure of the Spearman correlation for Phase 1. Fundamental and formant frequencies have statistically significant correlations with many variables.}
\end{figure*}

The correlation matrix of all variables is in \autoref{fig:p1correlation}.

\subsubsection{Study (2): Automated Code-based Manipulations}
Considering the descriptive statistics (\autoref{tab:p2descriptives}) with the interpretation suggested by \citet{seaborn_can_2023}, three voices were perceived as kawaii: the young boy けんしん (Kenshin) via Cubase-VA5 ($M=3.5, SD=0.8, MD=4, IQR=1$), STRAIGHT-VA5 ($M=3.5, SD=0.8, MD=4, IQR=1$), and WORLD-VA5 ($M=3.6, SD=0.8, MD=4.0, IQR=1$). In conflict with Phase 1, (SAYO$\beta$) was not perceived as kawaii through Cubase-VA2 ($M=3.3, SD=0.8, MD=3, IQR=1$), STRAIGHT-VA2 ($M=3.3, SD=0.9, MD=4, IQR=1$), and WORLD-VA2 ($M=3.2, SD=0.8, MD=3, IQR=1$). 


\begin{table*}[!ht]
    \caption{Descriptive statistics for VA kawaii perceptions by manipulation (Phase 2). $\triangle$: Kawaii.}
    \label{tab:p2descriptives}
    \centering
\begin{tabular}{l lllll}
\toprule
& \textbf{M} & \textbf{SD} & \textbf{MD} & \textbf{IQR} & \textbf{Kawaii} \\ \midrule
Cubase VA1 (Asano Yuki) & 3.3 & 0.9 & 4.0 & 1   & \\
Cubase VA2 (Sayo) & 3.3 & 0.8 & 3.0 & 1   & \\
Cubase VA3 (MAKOTO grandma) & 2.1 & 0.9 & 2.0 & 1.75 & \\
Cubase VA4 (MAKOTO grandpa) & 1.8 & 0.7 & 2.0 & 1   & \\
Cubase VA5 (Kenshin) & 3.5 & 0.8 & 4.0 & 1   & $\triangle$
 \\
\midrule 
Legacy-STRAIGHT VA1 (Asano Yuki) & 3.4 & 0.9 & 4.0 & 1   & \\
Legacy-STRAIGHT VA2 (Sayo) & 3.3 & 0.9 & 4.0 & 1   & \\
Legacy-STRAIGHT VA3 (MAKOTO grandma) & 2.0 & 0.9 & 2.0 & 1   & \\
Legacy-STRAIGHT VA4 (MAKOTO grandpa) & 1.7 & 0.7 & 2.0 & 1   & \\
Legacy-STRAIGHT VA5 (Kenshin) & 3.5 & 0.8 & 4.0 & 1   & $\triangle$
 \\
\midrule 
WORLD VA1 (Asano Yuki)  & 3.4 & 0.9 & 3.5 & 1   & \\
WORLD VA2 (Sayo)  & 3.2 & 0.8 & 3.0 & 1   & \\
WORLD VA3 (MAKOTO grandma)  & 2.0 & 0.8 & 2.0 & 2   & \\
WORLD VA4 (MAKOTO grandpa)  & 1.8 & 0.8 & 2.0 & 1   & \\
WORLD VA5 (Kenshin)  & 3.6 & 0.8 & 4.0 & 1   & $\triangle$
 \\ \bottomrule
\end{tabular}
\end{table*}

Next, we tested each hypothesis. 
Wilcoxon signed-rank tests between Cubase and the automatic speech processing methods showed significant differences in certain variables for some voices, summarized in \autoref{tab:p2results}.


\begin{table*}[!ht]
\caption{Wilcoxon signed-rank test results comparing Cubase to the automatic speech processing methods (Phase 2).}
\label{tab:p2results}
\centering
\begin{tabular}{lll}
\toprule
\textbf{Voice} & \textbf{Cubase vs Legacy-STRAIGHT}& \textbf{Cubase vs WORLD}\\
\midrule

VA1 Asano Yuki   & $t(50)=82.0, p=.59$& $t(50)=77.5, p=.45$\\

VA2 Sayo & $t(50)=70.5, p=.49$& $t(50)=68.0, p=.65$\\

VA3 MAKOTO grandma & $t(50)=45.5, p=.64$& $t(50)=80.5, p=.81$\\

VA4 MAKOTO grandpa & $t(50)=40.0, p=.20$& $t(50)=18.0, p=1.0$\\

VA5 Kenshin & $t(50)=91.0, p=.86$& $t(50)=95.0, p=.68$\\
\bottomrule
\end{tabular}
\end{table*}

We then carried out exploratory analyses of the non-kawaii variables, as a means of checking the extent to which each method produced the same overall audio manipulation.
However, Wilcoxon signed-rank tests showed significant differences in certain variables for some voices, as shown in \autoref{tab:p2other}. 

No other statistically significant differences were found.


\begin{table*}[!ht]
\centering
\caption{Wilcoxon signed-rank test results for other variables (Phase 2). Sig.: Statistically significant. *$p < .05$, **$p < .01$, ***$p < .001$.}
\label{tab:p2other}

\begin{tabular}{p{1cm}lllllll}
\toprule
\textbf{Variable} & \textbf{Voice} & \textbf{Cubase} & \bfseries\makecell[c]{Legacy-\\STRAIGHT} & \textbf{WORLD} & \textbf{$t$} & \textbf{$p$} & Sig. \\ \midrule
Human-like   &  Asano Yuki (CoeFont) & $M=2.4, SD=1.0$ & $M=2.8,SD=1.1$  & & $73.0$ & .02 & * \\ \midrule
\multirow[t]{3}{*}[-6pt]{\makecell[l]{Animal-\\like}}  &  Asano Yuki (CoeFont) & $M=1.9, SD=0.8$ & $M=2.1,SD=1.0$  & & $32.5$ & .02 & * \\
&  MAKOTO grandma & $M=2.1, SD=0.9$ & $M=2.3, SD=0.9$ & & $52.5$ & .02 & * \\
&  MAKOTO grandma & & $M=2.3, SD=0.9$ & $M=2.1, SD=0.9$ & $29.5$ & .04 & * \\ \midrule
Trust-worthy &  MAKOTO grandma & $M=2.5, SD=0.9$ & & $M=2.8, SD=0.7$ & $40.0$ & $<.001$ & *** \\ \midrule
\multirow[t]{3}{*}[-6pt]{\makecell[l]{Favour-\\able}}  &  MAKOTO grandpa & $M=2.6, SD=0.8$ & $M=2.9, SD=0.9$ & & $48.0$ & .02 & * \\
 &  MAKOTO grandpa & $M=2.6, SD=0.8$ & & $M=2.9, SD=0.8$ & $60.0$ & .02 & * \\ \midrule
Excited     &  Sayo (CoeFont) & $M=2.2, SD=0.9$ & $M=1.9, SD=0.6$ & & $47.5$ & .01 & * \\
 &  MAKOTO grandma & $M=1.6, SD=0.7$ & & $M=1.7, SD=0.6$ & $22.5$ & .03 & * \\ 
\bottomrule
\end{tabular}%
\end{table*}

\subsection{Discussion}
\label{sec:p1discussion}

Kawaii perceptions were correlated with some voice features, mainly high fundamental frequencies ($F0$), first formant frequencies ($F1$), and second formant frequencies ($F2$). Since $F1$ and $F2$ are believed to influence vowel determination~\cite{kent_static_2018,sadao2014formant,Kasuya1968jpformant}, this result is natural. However, gender perceptions~\cite{Kasuya1968jpformant,kent_static_2018} were not meaningfully correlated with perceptions of kawaiiness. This was unexpected, given the stereotype of kawaii as ``girlish''~\cite{shiokawa1999cute,yomota2006_kawaiiron} and the initial results for kawaii vocalics~\cite{seaborn_can_2023, seaborn_game_2023}.
The influence of formant frequencies may not be as great as that of $F0$~\cite{Hardy_2020}. \citet{Hardy_2020}, for instance, found that increasing $F0$ tended to lead to more feminine and gender ambiguous attributions.
Perhaps nonlinear effects or confounds were present; future work on a range of vocalics can evaluate this. 

Voice clips with higher $F3$ were perceived as more gender ambiguous, but this was not linked to escalating levels of perceived kawaiiness. Although $F3$ can affect perception of the difference between $r$ and $l$ pronunciations~\cite{dalston_acoustic}, minor pronunciation differences may not affect kawaii. There could be a ceiling effect, voice-specific ``sweet spots,'' or other vocalics that explain the earlier results of gender ambiguity and kawaii~\cite{seaborn_can_2023,seaborn_game_2023}. Also, voice clips with higher fundamental, $F1$, and $F2$ frequencies were perceived as younger, verifying the kawaii vocalics connection with age~\cite{seaborn_can_2023}. \citet{seaborn_can_2023} first hypothesized against an age effect based on the notion of \citet{lieber2021otona} otona-kawaii (adult-kawaii) and kawaii elicited by smiling older adults~\cite{nittono_psychophysiological_2017}. However, we found that kawaii favours youth, i.e., a young age effect.
Indeed, we were able to manipulate kawaii perceptions of these TTS voices with our procedure. 
Still, the unknown factors and unprecedented results may affect kawaii vocalics manipulations of other voices. Participants may be persuaded 
by positive- and neutral-toned VAs, as well as middle-aged masculine or younger feminine voices~\cite{pias2024impact}. In kind, users should be able to customize the gender, age, and kawaii levels of their VAs. 

The results for Study (2) showed that we were able to automate the manual procedure using two speech signal processing methods, with caveats.
Our manipulation of fundamental and formant frequencies with Cubase (DAW), Legacy-STRAIGHT, and WORLD  had the same or similar effects for the TTS voices. 
Yet, important differences were found for several variables related to but distinct from kawaii perceptions. Notably, 小夜 (SAYO$\beta$) was not as kawaii as in (1). 
This may be due to subtle differences in how each procedure estimates frequencies. In Cubase, we used Steinberg's VariAudio~\cite{cho2017computer}. For the code-based methods, we used estimation functions. In Legacy-STRAIGHT, we used MulticueF0v14 to extract fundamental frequencies and extrastraightspec to spectrogram information was extracted~\cite{kawahara1997straight}.
In WORLD, Dio, StoneMask, and CheapTrick were used~\cite{MasanoriMORISE20162015EDP7457_WORLD}.
Each differs in sound quality of the synthesized speech and processing speed~\cite{MasanoriMORISE20162015EDP7457_WORLD}.
Results may not be exactly the same when resynthesizing due to $F0$ estimation errors. We must be cautious about the results 
given these differences. We recommend using manipulation checks and pilot testing manipulations before use
~\cite{seaborn2024unbox}.
Future work may seek to refine the automated procedures through technical enhancements to each library or by creating a new library. Still, the most important results for kawaii held.
As such, we proceeded to apply a code-based kawaii vocalics manipulation method to a wider variety of voices.

\section{Phase 2: Application of Speech Signal Processing to Game Character Voices}
\label{sec:phase3}

The goal of this phase was to test whether the manipulation would work for a different sample of voices. Notably, we focused on prerecorded voices instead of generated TTS voices.
Therefore, we hypothesized:
\begin{quote}
    \textbf{H3:} A three-semitone shift in fundamental and formant frequencies will increase perceptions of kawaiiness  in game character voices.
\end{quote}

We used the dataset by \citet{seaborn_game_2023} to compare our manipulated voices to a baseline, focusing on kawaiiness and kawaii-linked  vocalics like humanlikeness~\cite{seaborn_can_2023,seaborn_game_2023}. We predicted the same results as for the TTS voices (\autoref{sec:phase1}).

\subsection{Materials}
We used the 18 game character voices from \citet{seaborn_game_2023}, which reflect different ages, genders, and vocal types. Game character voices are also different from VA TTS voices in terms of quality and expression, since the purpose is different (entertainment vs. service) and game character voices are prerecorded rather than generated. 
Given the results of Phase 1 (\autoref{sec:phase1}), we used WORLD synthesizers, a WORLD function, to manipulate all voices. The fundamental and formant frequencies were set to $2^{\frac{3}{12}}$ times higher than the original voices.

\subsection{Participants and Recruitment}
Participants ($N=150$, women $n=71$, men $n=78$, another gender or N/A $n=1$) through Yahoo! Crowdsourcing Japan on July 16\textsuperscript{th}, 2024. Most respondents were aged 45-54 ($n=56$) and 55-64 ($n=33$), with some younger (35-44, $n=26$) and older (65+, $n=8$).
Participants were compensated $\sim$600 yen for 30 minutes as per ethics guidelines.
Participants in other phases were excluded using ID identification to prevent participation in multiple studies.

\subsection{Data Analysis}
Descriptive statistics were calculated. Shapiro-Wilk tests indicated non-normal distributions, so the Mann-Whitney U test was used to compare perceived kawaiiness between the manipulated kawaii voices and the non-manipulated voices data from \citet{seaborn_game_2023}.
Next, a Spearman correlation test~\cite{spearman} was conducted to examine relationships among the variables. Lastly, a Generalized Estimating Equations (GEE) analysis~\cite{gee} was performed to determine whether known covariates, such as humanlikeness~\cite{seaborn_can_2023,seaborn_game_2023}, and potential covariates, such as trustworthiness, based on the correlations 
that significantly contributed to perceived kawaiiness.

\subsection{Results}
\label{sec:p3results}


Descriptive statistics are presented in \autoref{tab:p3voices}.

\begin{table*}[!ht]
    \centering
    \caption{Descriptive statistics for perceptions of kawaii, age, and gender in game character voice, ordered by kawaii rating (Phase 2).}
    \label{tab:p3voices}
    \resizebox{\textwidth}{!}{%
    \begin{tabular}{lllll}
    \toprule
        \textbf{Character} & \textbf{Game or Series} & \textbf{Kawaiiness} & \textbf{Age Group} & \textbf{Gender Group} \\ \midrule
        Barbara & Genshin Impact & $M=4.1, SD=0.9,MD=4, IQR=1$& Child (MD=2, 60\%) & Fem. (MD=2, 97\%) \\ \midrule
        Pikachu & The Pok\'{e}mon series & $M=3.9, SD=0.9,MD=4, IQR=0$& Child (MD=2, 50\%) & Amb. (MD=2, 97\%) \\ \midrule
        Edea & Bravely Default & $M=3.6, SD=0.8,MD=4, IQR=1$& Teen (MD=3, 48\%) & Fem. (MD=2, 87\%) \\ \midrule
        Ayaka & Genshin Impact & $M=4, SD=0.8,MD=4, IQR=0$& Adult (MD=4, 87\%) & Fem. (MD=2, 99\%) \\ \midrule
        QiQi & Genshin Impact & $M=3.7, SD=0.9,MD=4, IQR=1$& Child (MD=2, 58\%)  & Fem. (MD=2, 94\%) \\ \midrule
        Peach & The Super Mario series & $M=3.7, SD=0.8,MD=4, IQR=1$& Teen (MD=3, 43\%)  & Fem. (MD=2, 92\%) \\ \midrule
        Kirby & The Kirby series & $M=3.6, SD=0.9,MD=4, IQR=1$& Child (MD=3, 69\%)  & Amb. (MD=3, 51\%) \\ \midrule
        Ashley & The Wario series & $M=3.2, SD=0.9,MD=3, IQR=1$& Teen (MD=3, 68\%) & Fem. (MD=2, 96\%) \\ \midrule
        Zelda & TLoZ series & $M=3.6, SD=0.8,MD=4, IQR=1$& Adult (MD=4, 63\%) & Fem. (MD=2, 98\%) \\ \midrule
        Jigglypuff & The Pok\'{e}mon series & $M=3, SD=1,MD=3, IQR=2$& Child (MD=2, 45\%) & Amb. (MD=3, 43\%) \\ \midrule
        Toadette & The Super Mario series & $M=3, SD=1,MD=3, IQR=2$& Child (MD=2, 59\%) & Fem. (MD=2, 55\%) \\ \midrule
        Yoshi & The Super Mario series & $M=3.4, SD=1,MD=3.5, IQR=1$& Baby (MD=2, 57\%) & Amb. (MD=3, 53\%) \\ \midrule
        Young Link & TLoZ series & $M=3.3, SD=0.9,MD=3, IQR=1$& Child (MD=2, 62\%) & Masc. (MD=1, 55\%) \\ \midrule
        Baby Bowser & The Super Mario series & $M=2.7, SD=1,MD=3, IQR=1$& Child (MD=2, 35\%) & Amb. (MD=3, 45\%) \\ \midrule
        Toad & The Super Mario series & $M=3, SD=0.9,MD=3, IQR=1$& Child (MD=2, 55\%) & Masc. (MD=1, 54\%) \\ \midrule
        Inkling Girl & Splatoon 3 & $M=2.6, SD=1,MD=3, IQR=1$& Child (MD=2, 41\%) & Amb. (MD=3, 41\%) \\ \midrule
        Luma & Super Mario Galaxy & $M=2.3, SD=0.9,MD=2, IQR=1$& Ageless (MD=7, 76\%) & Neu. (MD=4, 69\%) \\ \midrule
        Shizue & Animal Crossing, e.g. & $M=2.8, SD=1,MD=3, IQR=2$& Ageless (MD=7, 70\%) & Neu. (MD=4, 57\%) \\ 
        \bottomrule
    \end{tabular}
    }
\end{table*}


\begin{figure*}[!ht]
    \centering
    \includegraphics[width=\textwidth]{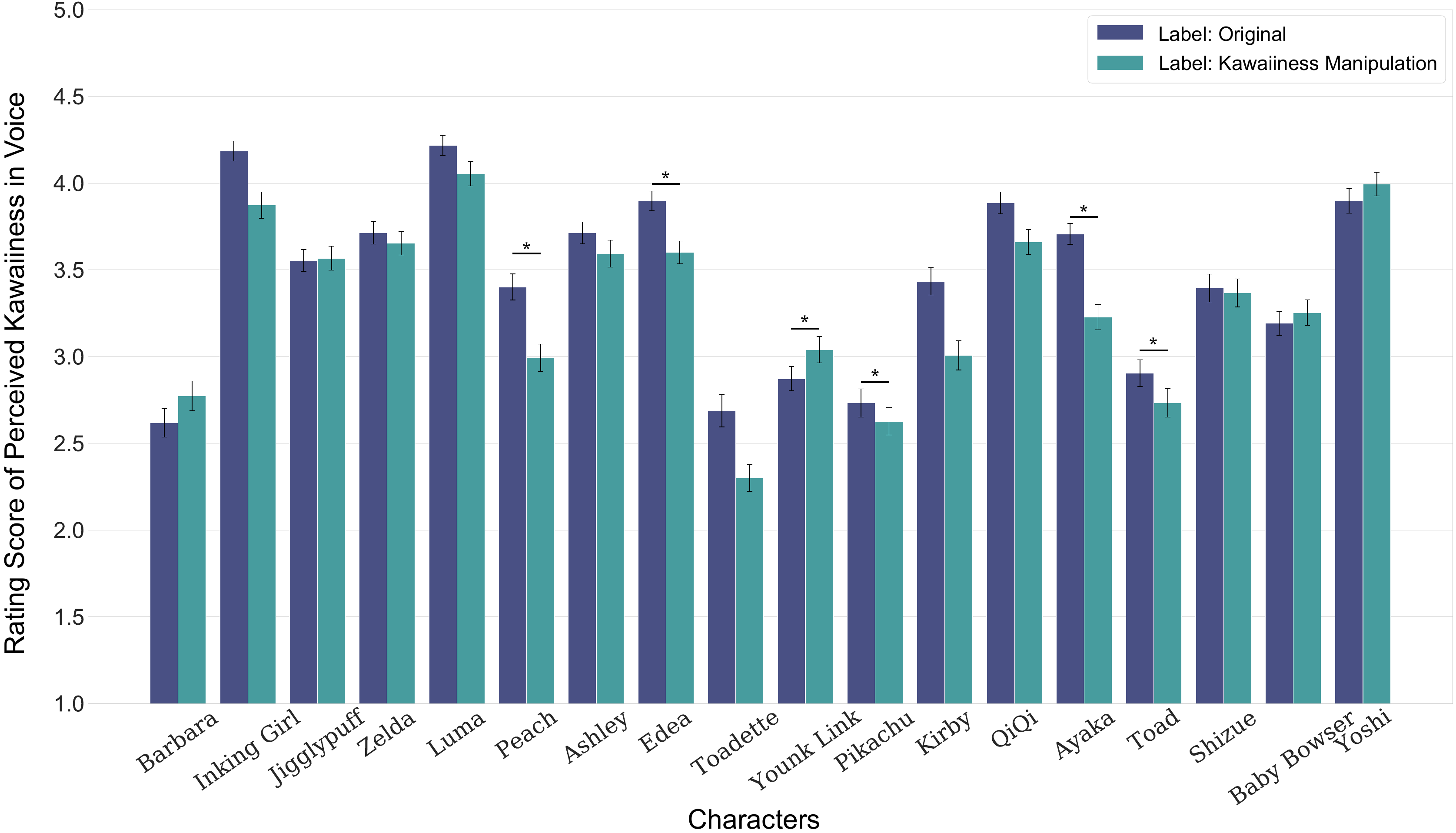}
    \caption{Mean distribution of perceived kawaiiness by game character for the original and kawaii-modified voices (Phase 2).}
    \label{fig:p3mean}
    \Description{Figure of the mean distribution of all characters for Phase 2.}
\end{figure*}


A Mann-Whitney U test revealed statistically significant differences ($U=4116959.50, N_{original}=157*18, N_{kawaii}=151*18, p<.001, r=.07$) 
in perceived kawaiiness between the manipulated voice ($M=3.30, SD=1.04$) and the original voices ($M=3.44, SD=1.02$).
The manipulation seemingly influenced perceived voice kawaii. The effect size was small ($r=.07$), possibly because of the relatively large sample size. Therefore, we cannot accept the hypothesis:

\begin{quote}
    \textbf{H3:} A three-semitone shift in fundamental and formant frequencies will increase perceptions of kawaiiness  in game character voices.\\ \textbf{$\rightarrow$ Rejected (for these game character voices).}
\end{quote}


\begin{table*}[!ht]
\centering
\caption{Mann-Whitney U test results comparing the original voices to the kawaii manipulated versions 
(Phase 2). *$p < .05$, **$p < .01$, ***$p < .001$. CI: Confidence Interval.}
\label{tab:s3resultsmann}

\renewcommand{\arraystretch}{1.0}
\def\arraystretch{1.05}%
\setlength{\tabcolsep}{6pt} 
\begin{tabular}{lllllll}
\toprule

\textbf{Voice} & \textbf{$U$} & \textbf{$Z$} & \textbf{$p$-value} & \textbf{$r$} & \textbf{$95\%CI$} & \textbf{Significance} \\ 
\midrule

Kawaii Manipulation vs. Original & 4116959.5 & 5.09 & $8.1e-08$ & 0.07 & [.04, .09] & *** \\ 

\bottomrule
\end{tabular}
\end{table*}

\begin{figure*}[!ht]
    \centering
    \includegraphics[width=\textwidth]{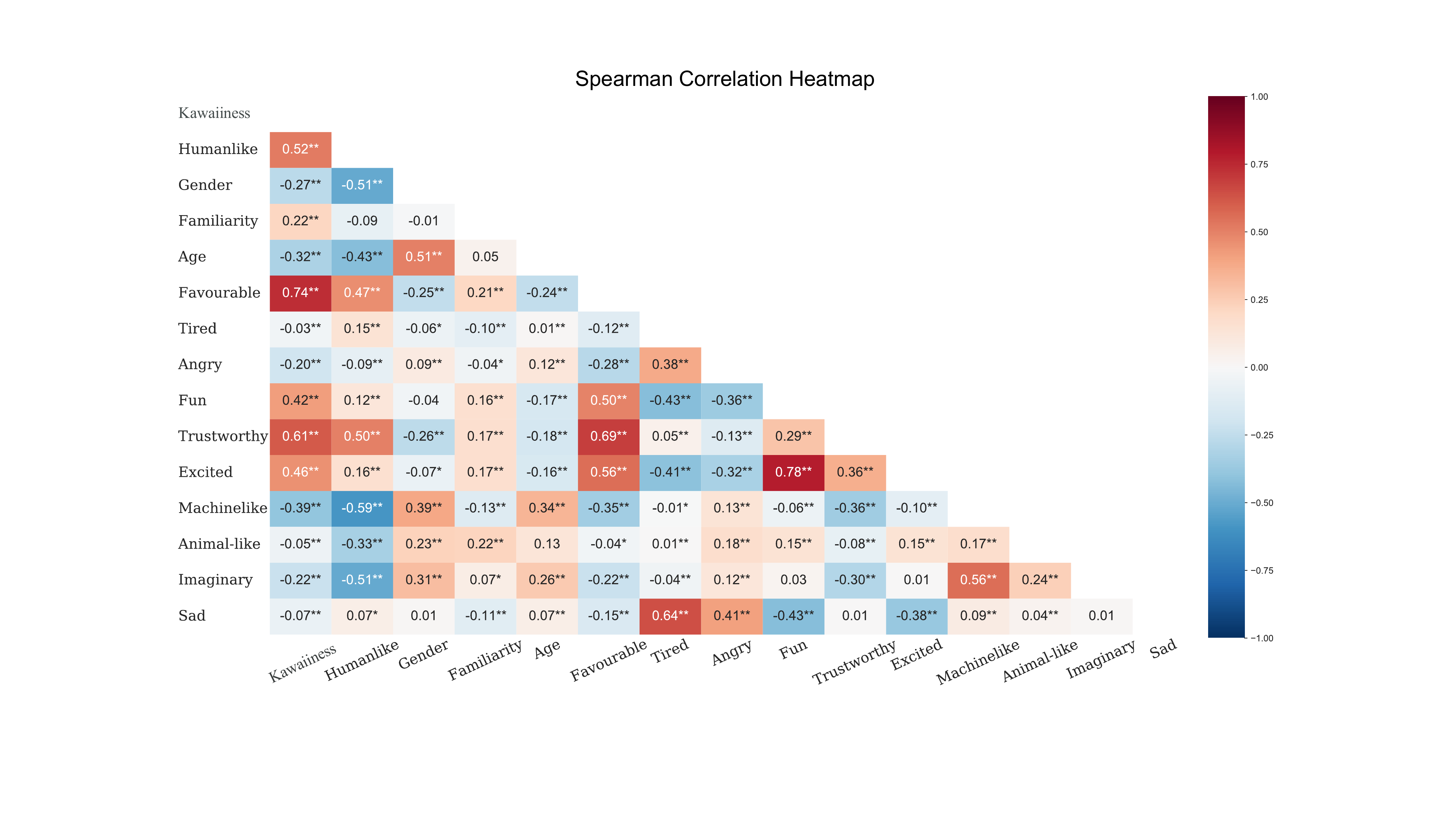}
    \caption{Spearman correlation test results for all variables (Phase 3). *$p < .05$, **$p < .01$.}
    \label{fig:p3correlation}
    \Description{Figure of the Spearman correlation for Phase 3.}
\end{figure*}

Spearman's tests (\autoref{fig:p3correlation}) found that several variables were significantly positively correlated with perceived kawaiiness: ``Favorable'' ($r_s=.74, p<.01$), ``Trustworthy'' ($r=.61, p<.01$), ``'Humanlike'' ($r=.52, p<.01$), ``Excited'' ($r=.46, p<.01$), ``Fun'' ($r=.42, p<.01$), and ``Familiarity'' ($r=.22, p<.01$), suggesting that higher ratings in these aspects were associated with higher perceived kawaiiness. Conversely, variables such as ``Machine-Like'' ($r=-.39, p<.01$), ``Age'' ($r=-.32, p<.01$), and ``Angry'' ($r=-.20, p<.01$) were significantly negatively correlated with perceived kawaiiness, indicating that higher ratings in these areas were associated with lower perceived kawaiiness.


The GEE analysis further investigated the influence of the highly correlated variables on perceived kawaiiness. It was found that ``Favourable'' ($coef=0.51, r=.16$), ``Humanlike'' ($coef=0.19, r=.06$), ``Familiarity'' ($coef=0.14, r=.04$), and ``Trustworthy'' ($coef=0.13, r=.04$) statistically significantly contributed to the perceived kawaiiness of the voice ($p<.001$). 
This suggests that these co-variables play an important role in shaping the perception of a voice's kawaiiness.


\begin{table*}[!ht]
\centering
\caption{GEE regression results for analyzing the co-variables contributing to voice Kawaiiness (Phase 3). *$p < .05$, **$p < .01$, ***$p < .001$. CI: Confidence Interval.}
\label{tab:s3resultsgee}

\renewcommand{\arraystretch}{1.0}
\def\arraystretch{1.05}%
\setlength{\tabcolsep}{10pt} 
\begin{tabular}{llllll}
\toprule

\textbf{Variable} & \textbf{$Coef$} & \textbf{$p$-value} & \textbf{$r$} & \textbf{$95\%CI$} & \textbf{Significance} \\ 
\midrule
Favourable & .51 & <.001 & .16 & [.14, .18] & *** \\ 
Human-like & .19 & <.001 & .06 & [.05, .07] & *** \\ 
Familiarity & .14 & <.001 & .04 & [.03, .05] & *** \\
Trustworthy & .13 & <.001 & .04 & [.02, .06] & *** \\ 
Excited & .09 & <.001 & .03 & [.02, .04] & *** \\ 

\bottomrule
\end{tabular}
\end{table*}

\subsection{Discussion}
\label{sec:p3disussion}

The results differed from Phase 1 (refer to \ref{sec:p1results}). 
Here, we used prerecorded game character voices instead of generated TTS voices. Voices created for game characters are typically recorded by voice actors in a professional studio and edited if necessary. Several had filters and audio manipulations like echoes and vibrato. 
The voices may have become uncanny, conflicting with the humanlikeness predicted by the kawaii vocalics model~\cite{seaborn_can_2023}.
This differences may also be linked to the estimation methods, as mentioned in\ref{sec:p1discussion}. In WORLD, aperiodicity of sound was estimated by D4C, but processing introduced noise for many character voices. 
Notably, Kirby and Baby Bowser may have sounded filtered, affected perceptions of fluency and naturalness and thereby decreased kawaii.

Given the results and their apparent grounding in the automation procedures, we returned to the confirmed manual method---Cubase---to verify the kawaii procedure from a user perception standpoint. We also considered the utility of finer-grained semitone manipulations---one, two, and three---in case some voices were sensitive to semitone adjustment.

\section{Phase 3: Granular Manipulation of  Game Character Voices}
\label{sec:phase4}

In the last phase, we aimed to replicate the results from the manual kawaii method applied to TTS voices in Phase 1 (\autoref{sec:phase1}) for game character voices. The results (refer to \ref{sec:p3results}) 
suggested a conflict between automating an extreme manipulation and highly processed and filtered voice samples. As such, we sought to confirm whether a kawaii amplification effect could still occur with the manual Cubase method, whereby we could manually account for auditory conflicts. Our hypothesis was similar but reflected this fine-grained approach to semitone adjustments: 
\begin{quote}
    \textbf{H4:} A one- or two-semitone shift in fundamental and formant frequencies will increase perceptions of kawaiiness  in game character voices.
\end{quote}

\subsection{Materials}
For finer-grained frequency manipulations, we made three versions of the fundamental and formant frequencies: one, two, and three semitones. 
In other words, we prepared voices that were manipulated $2^{\frac{1}{12}}$, $2^{\frac{2}{12}}$, and $2^{\frac{3}{12}}$ times higher than the original voices.  
As in Phase 1, all manipulations were performed in Cubase.

\subsection{Participants and Recruitment}
Participants ($N=51$, women $n=8$, men $n=43$, none of another gender identity) through Yahoo! Crowdsourcing Japan on August 22\textsuperscript{th}, 2024. Respondents were aged 45-54 ($n=18$), 55-64 ($n=10$), 35-44 ($n=11$), and 65+ ($n=5$).
As before, participants were compensated $\sim$600 yen for 30 minutes as per ethics guidelines.
Participants in other phases were excluded using ID identification to prevent them from participating in multiple studies.

\subsection{Data Analysis}
Descriptive statistics were created.
Given that the data was not normally distributed (via the Shapiro-Wilk test), the nonparametric Friedman~\cite{friedman_test} and Wilcoxon signed-rank~\cite{wilcoxon_test} tests were used.
Friedman tests were conducted to determine whether there were statistically significant differences in voice ratings among the different semitones per game character voice and across all characters combined. 
Post hoc pairwise comparisons were performed using Wilcoxon signed-rank tests to identify specific differences between pairs of semitones by character and across all characters. 
Also, as in Phase 1, representative values for the fundamental and formant frequencies of each voice were also created. Plots were generated to show the relationship between kawaii and each variable.

\subsection{Results}
\label{sec:p4results}

\begin{figure*}[!ht]
    \centering
    \includegraphics[width=\textwidth]{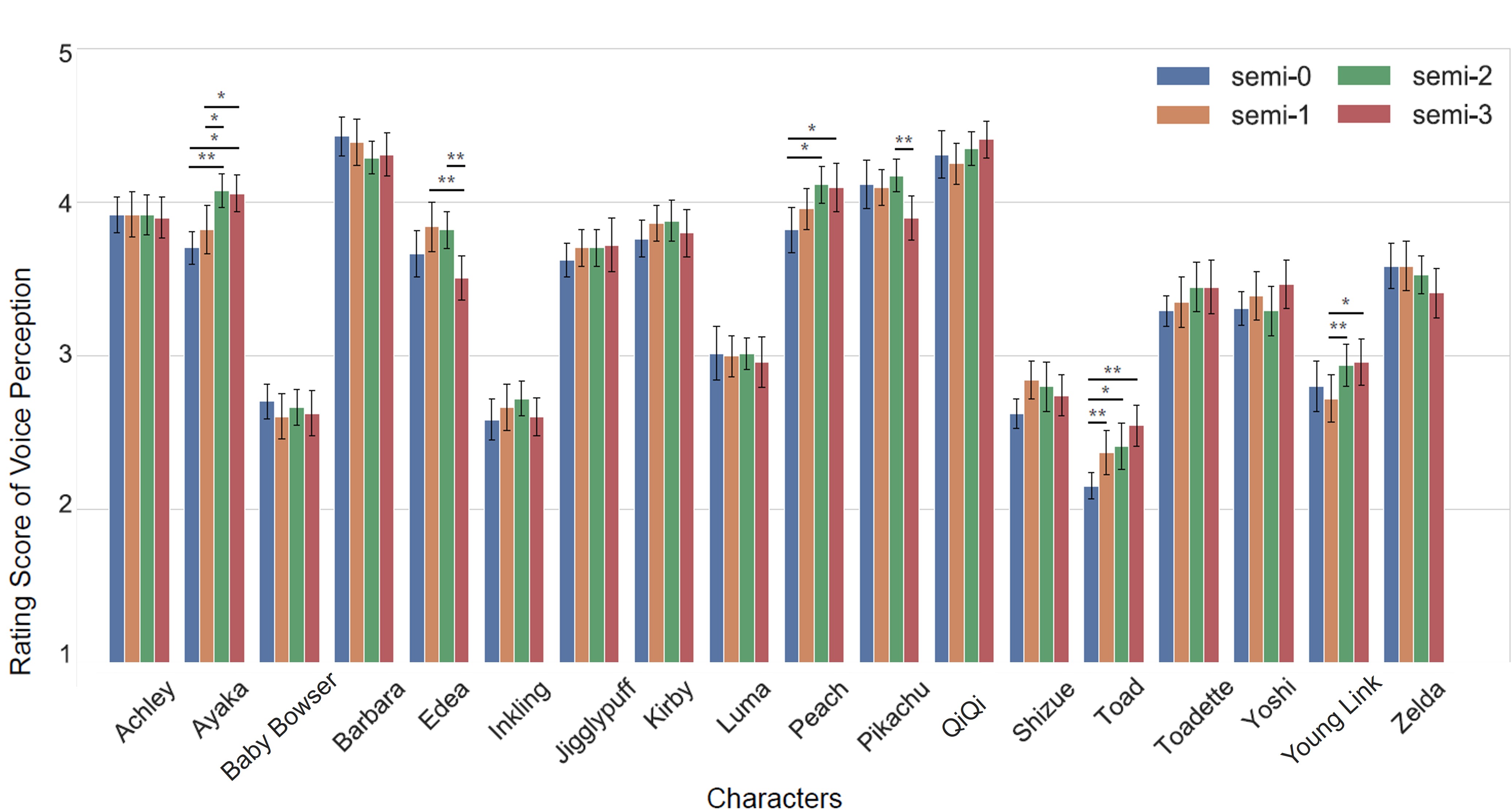}
    \caption{Mean perceived kawaiiness by voice and semitone (Phase 3).}
    \label{fig:p4mean}
    \Description{Figure of the mean kawaiiness for each character and  semitone, showing that overall kawaii ratings varied and semitone variance depended on the voice.}
\end{figure*}

\autoref{fig:p4mean} shows kawaiiness mean of each voice by semitone (semi-0, 1, 2, 3 correspond to the original voice and those manipulated to match the number of semitones). Descriptive statistics are provided in the Supplementary Materials.

Spearman's correlation analyses (\autoref{fig:p4correlation}) found negative and statistically significant relationships between kawaiiness and fundamental, first, second, and third formant frequencies. The increase in kawaii seen in the TTS was not seen in the game characters.

\begin{table}[!ht]
\caption{Spearman's correlations between perceived kawaii and fundamental and formant frequencies (Phase 3).}
\label{fig:p4correlation}
\begin{tabular}{l|lll}
\toprule
\textbf{Correlation} & {\color[HTML]{1F1F1F} \textbf{$r_s(70)$}} & \textbf{$p$-value} & {\color[HTML]{1F1F1F} \textbf{95\% $CI$}} \\ \midrule
Kawaii vs $F0$ (mel) & $-0.58$ & $p < .001$ & {\color[HTML]{212529} $[-0.73, -0.38]$} \\
Kawaii vs $F1$ (mel) & $-0.69$ & $p < .001$ & {\color[HTML]{212529} $[-0.81, -0.52]$} \\
Kawaii vs $F2$ (mel) & $-0.33$ & $p =.004$  & {\color[HTML]{212529} $[-0.53, -0.10]$} \\
Kawaii vs $F3$ (mel) & $-0.49$ & $p < .001$ & {\color[HTML]{212529} $[-0.66, -0.27]$} \\ 
\bottomrule
\end{tabular}
\end{table}

The results of the Friedman test (\autoref{tab:s4resultsbychara}) revealed statistically significant differences in voice ratings among the semitones for several characters and across all characters ($X^2=18.25, p<.001, W=8436.04$).
Specifically, significant differences were observed for ``Edea'' ($X^2=10.81, p=.013, W=.03$), ``Pikachu'' ($X^2==8.99, p=.030, W=.02$), ``Toad'' ($X^2=17.96, p<.001, W=.04$), ``Young Link'' ($X^2=8.40, p=.039, W=.02$), ``Ayaka'' ($X^2=15.84, p=.001, W=.04$), and ``Peach'' ($X^2=9.08, p<.028, W=.02$). 
Other characters did not show significant differences among the semitones, indicating that voice manipulation had varied effects on voice rating, but depends on the character.

\begin{table*}[!ht]
\centering
\caption{Statistically significant Wilcoxon signed-rank tests as post hoc pairwise comparisons for semitone manipulation by game character (Phase 3). *$p<.05$, **$p<.01$. CI: Confidence Interval. Direction: $++$ Perceived kawaiiness increased. $--$ Perceived kawaiiness reduced. $==$ No change in perceived kawaiiness.}
\label{tab:s4resultsposthoc}
\renewcommand{\arraystretch}{1.0}
\def\arraystretch{1.01}%
\setlength{\tabcolsep}{6pt} 
\begin{tabular}{lllllll}
\toprule

\textbf{Semitones} & \textbf{Character} & \textbf{Direction} & \textbf{$W$} & \textbf{$p$-value} & \textbf{$r$} & \textbf{Significance} \\ \midrule

\multirow[t]{2}{*}{semi-0 vs. semi-1} 
& All & $++$ & 21500.00 & .037 & 23.42 & * \\ 
& Toad & $++$ & 27.00 & .008 & 0.53 & ** \\ 
\midrule

\multirow[t]{4}{*}{semi-0 vs. semi-2} 
& All & $++$ & 29493.00 & .000 & 32.13 & ** \\ 
& Toad & $++$ & 27.00 & .026 & 0.53 & * \\ 
& Ayaka & $++$ & 78.00 & .002 & 1.53 & ** \\ 
& Peach & $++$ & 84.00 & .011 & 1.65 & * \\ \midrule

\multirow[t]{4}{*}{semi-0 vs. semi-3} 
& All & $++$ & 41558.50 & .040 & 45.27 & * \\ 
& Toad & $++$ & 28.50 & .001 & 0.56 & ** \\ 
& Ayaka & $++$ & 181.50 & .020 & 3.56 & * \\ 
& Peach & $++$ & 112.50 & .025 & 2.21 & * \\ \midrule

\multirow[t]{3}{*}{semi-1 vs. semi-2} 
& All & $++$ & 16496.50 & .050 & 17.97 & * \\ 
& Young Link & $++$ & 14.00 & .008 & 0.27 & ** \\ 
& Ayaka & $++$ & 45.00 & .016 & 0.88 & * \\ \midrule

\multirow[t]{4}{*}{semi-1 vs. semi-3} 
& All & $==$ & 23961.50 & .840 & 26.10 & \\ 
& Edea & $--$ & 50.00 & .004 & 0.98 & ** \\ 
& Young Link & $++$ & 32.00 & .022 & 0.63 & * \\ 
& Ayaka & $++$ & 96.00 & .048 & 1.88 & * \\ \midrule

\multirow[t]{3}{*}{semi-2 vs. semi-3} 
& All & $--$ & 16750.50 & .101 & 18.25 & \\ 
& Edea & $--$ & 26.50 & .003 & 0.52 & ** \\ 
& Pikachu & $--$ & 24.00 & .007 & 0.47 & ** \\ 
\bottomrule

\end{tabular}
\end{table*}

The post-hoc analyses further elucidated the differences between pairs of semitones within the characters and across all characters (\autoref{tab:s4resultsposthoc}). 
First, the significant difference across all character was observed for the pairs of semitones: semitone-0 and semitone-1 ($W=21500.00, p=.037, r=23.42$), semitone-0 and semitone-2 ($W=29493.00, p<.001, r=32.13$), semitone-0 and semitone-3 ($W=41558.50, p=.040, r=45.27$), semitone-1 and semitone-2 ($W=16496.50, p=.050, r=17.97$).

\begin{table*}[!ht]
\centering
\caption{Statistically significant Friedman test results for semitone manipulation by game character (Phase 3). *$p < .05$, **$p < .01$, ***$p < .001$. CI: Confidence Interval.}
\label{tab:s4resultsbychara}
\renewcommand{\arraystretch}{1.0}
\def\arraystretch{1.05}%
\setlength{\tabcolsep}{6pt} 
\begin{tabular}{llllll}
\toprule

\textbf{Character} & \textbf{$X^2$} & \textbf{$p$-value} & \textbf{$W$} & \textbf{95\% CI} & \textbf{Significance} \\ 
\midrule

All & 18.25 & $< .001$ & 8436.04 & [8202.54, 8660.19] & *** \\ 

Edea & 10.81 & .013 & .03 & [0.58, 1.31] & * \\ 
Pikachu & 8.99 & .030 & .02 & [1.12, 1.90] & * \\ 
Toad & 17.96 & $< .001$ & .04 & [0.02, 0.13] & *** \\ 
Young Link & 8.40 & .039 & .02 & [0.03, 0.24] & * \\ 
Ayaka & 15.84 & .001 & .04 & [0.91, 1.62] & ** \\ 
Peach & 9.08 & .028 & .02 & [1.05, 1.72] & * \\ 
\bottomrule
\end{tabular}
\end{table*}

Significant differences were found by character between semitone-0 and semitone-1 for the character ``Toad'' ($W=27.00, p=.008, r=.53$). 
Differences between semitone-0 and semitone-2 were significant for ``Toad'' ($W=27.00, p=.026, r=.53$), ``Peach'' ($W=84.00, p=.011, r=1.65$) and ``Ayaka'' ($W=78.00, p=.002, r=1.53$). 
Differences between semitone-0 and semitone-3 were significant for ``Toad''  ($W=28.50, p=.001, r=.56$), ``Peach'' ($W=112.50, p=.025, r=2.21$), and ``Ayaka'' ($W=181.50, p=.020, r=3.56$). 
Other significant comparisons included 
semitone-1 vs. semitone-2 for ``Young Link'' ($W=14.00, p=.008, r=.27$) and ``Ayaka'' ($W=45.00, p=.016, r=.88$), 
semitone-1 vs. semitone-3 for ``Edea'' ($W=50.00, p=.004, r=.98$), ``Young Link'' ($W=32.00, p=.022, r=.63$) and ``Ayaka'' ($W=96.00, p=.048, r=1.88$), and 
semitone-2 vs. semitone-3 for ``Edea'' ($W=26.50, p=.003, r=.52$) and ``Pikachu'' ($W=24.00, p=.007, r=.47$), 
indicating that specific manipulations of voice had distinct impacts on voice rating depending on the semitone level and the character involved.
Refer to \autoref{tab:s4resultsposthoc} for direction (kawaii$++$ or kawaii$--$).

In sum, we can partially accept the hypothesis:
\begin{quote}
    \textbf{H4:} A one- or two-semitone shift in fundamental and formant frequencies will increase perceptions of kawaiiness  in game character voices.
    \\ \textbf{$\rightarrow$ Partially accepted (for certain game character voices).}
\end{quote}

\subsection{Discussion}
\label{sec:p4discussion}

\begin{figure*}[!ht]
    \centering
    \includegraphics[width=.85\textwidth]{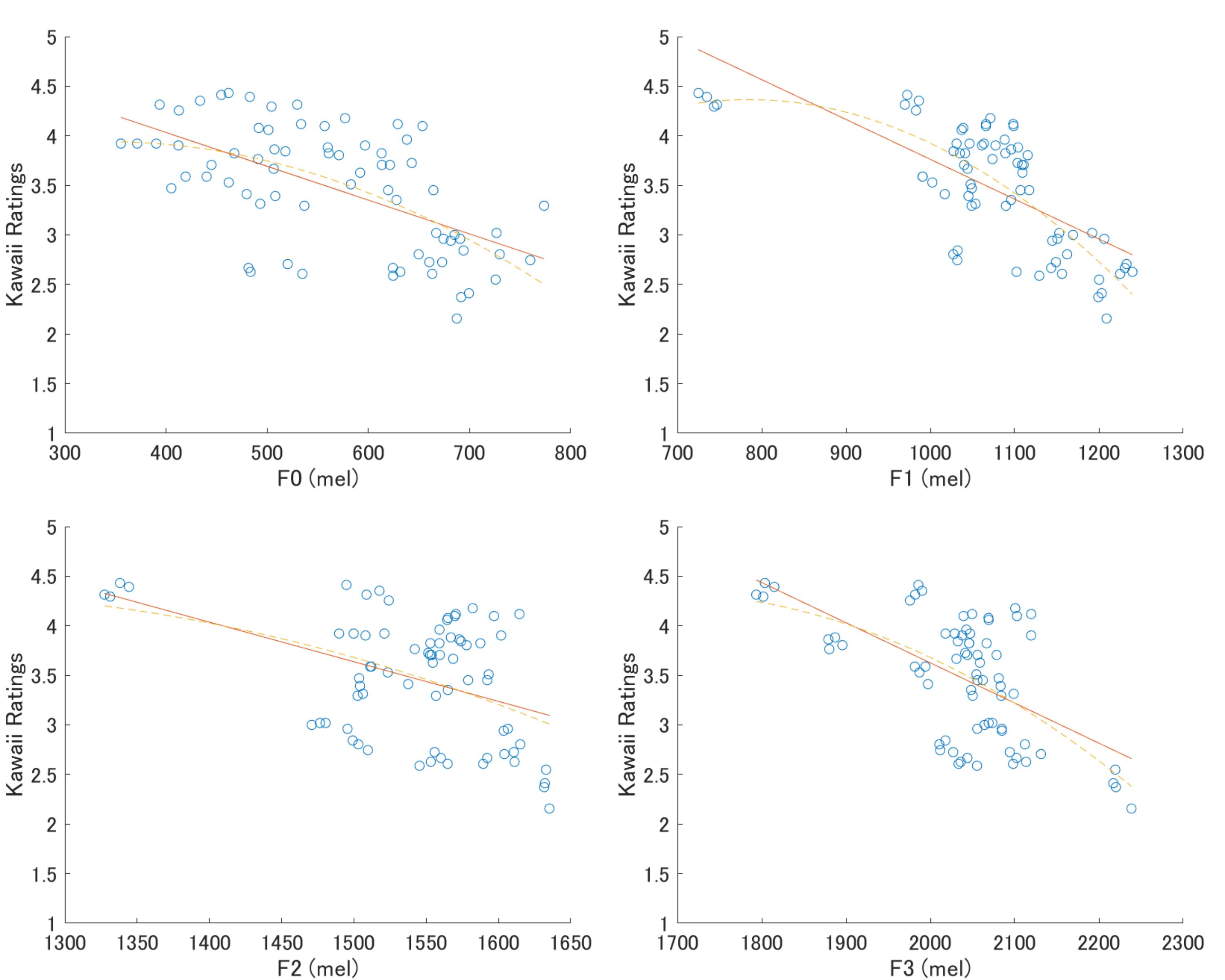}
    \caption{Scatter plots by kawaii perceptions and $F0$, $F1$, $F2$, $F3$ (Phase 3).}
    \label{fig:p4scatter}
    \Description{Scatter plots by kawaii perceptions and $F0$, $F1$, $F2$, and $F3$. There is a small negative correlation, but no sweet point was found.}
\end{figure*}

We were able to manipulate kawaii perceptions of certain voices in a statistically detectable way via manipulation of fundamental and formant frequencies. However, the direction of the manipulation depended on the voice. 
Although the voice analysis captured the overall trend, our frequency manipulations  did not always result in a linear increase in kawaiiness, as in Phase 1 (\autoref{sec:phase1}). The approximate line on the scatter plot (\autoref{fig:p4scatter}) showed a vertex in the quadratic approximation of $F1$ with kawaii perceptions. Still, confirming 
that there were no samples with 
$F1$ in the vicinity remains elusive. We also did not find manipulable ``sweet spots'' 
and a linear decrease was observed. We discuss the implications of these results in our overall discussion next.

\section{Overall Discussion}
\label{sec:discussion}

We attempted to enhance perceptions of voice kawaiiness via manipulation of fundamental and formant frequencies for TTS and game voices. TTS voices could be statistically significantly enhanced, but conversely, declines were observed for some game voices. Ultimately, we offer evidence for and a means of \emph{modifying} the kawaii features of voices, i.e., \textbf{
kawaii±}, that in certain cases can \emph{amplify} kawaii vocalics, i.e., \textbf{kawaii++}. 

In answer to \textbf{RQ1}, we found that
the fundamental and formant frequencies were voice features that could be manipulated to induce perceptions of voice kawaii. Specifically, we found that manipulation of these frequencies could amplify kawaii for certain voices in a certain range. This was especially true for TTS voices. However, the same manipulation could also subdue perceptions of kawaiiness  in some cases. We were not able to identify what specific features of the voices contributed to kawaii++ or kawaii--. Future work will need to explore these voices in more detail or consider other voice or sound features that may have confounded the results.

Our findings in Phases 1 and 3 indicated that the social identity factors of gender and age intersected with kawaii perceptions (\textbf{RQ2}). For the TTS voices, age perceptions were based on $F1$ and $F2$. Yet, the fundamental and all formant frequencies did not lead to greater perceptions of gender ambiguity, as predicted from the kawaii vocalics model~\cite{seaborn_can_2023}. More work is needed to tease out the reasons behind this. Perhaps there is a ceiling effect for manipulating these frequencies when it comes to gender and age ambiguity. 

We discuss specific findings and implications in detail next.

\subsection{Artificial Voices are More Kawaii? Current Technological Limitations of TTS Voices}
As shown in Phase 1, manipulation of certain frequencies improved the kawaii perceptions of TTS voices. This result contrasts with those of the Phase 3 for game character voices, which were prerecorded from human voices or otherwise crafted using professional techniques. This may indicate that TTS voices have room for further improvement compared to prerecorded voices in terms of factors like humanlikeness and fluency linked to kawaii~\cite{seaborn_can_2023}.
In fact, \citet{cambre2020inproceedings} found that real human voices still significantly outperform TTS voices in overall quality ratings, intelligibility, and speech speed. Similarly, generated TTS voices, by their nature, may always be limited in terms of kawaiiness compared to human voices. As per the kawaii vocalics model~\cite{seaborn_can_2023}, humanlikeness is a key factor. Still, when a low-to-medium quality TTS voice needs to be more kawaii, the approach in this study could be generally effective. 

\subsection{Voice Acting vs. Kawaii: The Need for Qualitative Insights}
The limits in our Phase 3 kawaii vocalics manipulations may be due to the prerecorded nature of the game character voices. Professional voice actors and studios may have maximized the level of kawaii for these voices. Still, there is no one type of ``kawaii'' voice. In professional and fan circles, kawaii has subtypes, and voice actors explore and enact diverse approaches to kawaii vocalics. For example, some have explored the concepts of otona-kawaii (adult-kawaii), in which older people are deemed visually kawaii by expressing positive emotions~\cite{nittono_psychophysiological_2017} or in personal expression, including appearance, fashion choices, and speech style~\cite{lieber2021otona}. 
As yet, a systematic study of these practices has yet to be done. 
This should be investigated through qualitative methods, such as interviews with voice actors and those around them, regarding the techniques that bring about voice kawaii. 
Such qualitative inquiry may also explain the underlying social factors that give rise to kawaii, e.g., when and why ``gender neutrality'' and ``gender ambiguity'' can be kawaii~\cite{seaborn_game_2023,seaborn_can_2023}. Another possible line of inquiry could be cross-cultural components; since kawaii is exported and taken up globally, it may be re-imagined and expressed---including verbally and vocally---in unique non-Japanese cultural ways~\cite{urakami2021cultural}. Future work can explore how voice actors can intentionally manipulate these concepts and why TTSs cannot.

We now outline an agenda for future work to further explore the procedures that have merit and confirm, refute, and/or extend the results so far:

\begin{itemize}
    \item \textbf{More fine-grained speech analysis:} Although we relied on averages, 
    we can test new hypotheses by focusing on time-series-specific data related to fundamental and formant frequencies. For example, by generating very short voice clips (less than one second) and breaking them down into vowels~\cite{Kasuya1968jpformant,igeta2011vowel}, consonants~\cite{Kumagai_2020}, and other pronunciations, it may be possible to conduct a more detailed investigation of kawaii perceptions.

    \item \textbf{Filtering by participant age:} No one under the age of 18 participated in this study. According to \citet{klatte_noise}, the effect of noise on listening comprehension differs between children and adults, so different results may be obtained across different age groups. People of different generations may also have different levels of exposure to kawaii voice phenomena or preferences that influenced perceptions in undetectable ways. Future work should explore cross-sections of participant age.
    
    \item \textbf{Cross-cultural verification:} The cross-cultural appeal of kawaii is well-established~\cite{Nittono2021crosscultural,Nittono2023crossculturalwords}. Yet, kawaii is still distinguished from other cultural notions of ``cute,'' to varying degrees~\cite{Nittono2023crossculturalwords,Nittono2023crossculturalwords}. 
    Future work should explore whether and how kawaii vocalics crosses sociocultural and sociolinguistic borders. A clear next step is to run user perception studies with non-Japanese populations. 
    
    \item \textbf{Cross-cultural mixing by region and language of listeners and voices:} Voices speaking in other languages or the use of Japanese voice clips but recruiting participants of varying regional origins may yield results that are not limited by cultural or regional differences. Notably, the formant frequencies of vowels differ from one language to another~\cite{Kasuya1968jpformant, zhenglai_analysis_2003}, so the manipulation of formants may lead to similar or different results.

    \item \textbf{Moving beyond survey instruments
    :} Now that the $F0$ and formant frequencies have been identified as specific variables that contribute to kawaii, kawaii vocalics manipulations can be validated in actual use cases. For example, it may be possible to train MCs  and voice actors to use this procedure, or conduct site-specific experiments where kawaii vocalics is relevant, like fan conventions and daycare centres.

    \item \textbf{Ethical issues in applying kawaii voices:} Use of kawaii voices may raise ethical issues. Kawaii may elicit certain coercive effects due to the link to baby schema~\cite{nittono_psychophysiological_2017,lorenz_angeborenen_1943}. If kawaii impressions are subconscious, the likelihood that impressions will be manipulated in direction desired by the voice provider will increase. This could lead users to make choices that were not originally desired. In other words, kawaii could be a dark pattern or deceptive design strategy~\cite{mathur2019darkpatterns,Brignull_2023}, an idea already explored for ``cute'' home robots~\cite{Lacey_2019}.
    
    \item \textbf{Impact on comprehensive voice UX metrics:} While several studies have attempted to develop overarching criteria for voice UX, no consensus has been reached~\cite{seaborn_measuring_2021}. Determining the role of specific phonetic features would be a useful outcome for the field and kawaii vocalics specifically. Qualitative methods like diary studies with participants paired over time with voices agents using voices of varying kawaii could be fruitful~\cite{seaborn2024qualvoice}. 
\end{itemize}

\subsection{Limitations}
\label{sec:limitations}

We did not conduct an analysis of speech content, which could influence results~\cite{baird_perception_2017}. We did not have a validated measure for kawaii, which remains elusive~\cite{wang2024kawaiicomp}. 
We could not eliminate any effect of time or repetition due to the large number of voice clips for each participant. 
As noted by \citet{beth_selection}, there are often unpredictable demographic biases in online surveys; in our case, we had age biases and did not include anyone under the age of 18.

\section{Conclusion}
Kawaii vocalics is a new field of study with great promise for HCI, HRI, and HAI. Here, we have shown that manipulating kawaii vocalics---manually or with automated methods---is possible, with caveats. As more artificial agents, interfaces, spaces, and systems are imbued with voices of all kinds, the importance of considering vocalics will grow.

\end{CJK}

\begin{acks}
This work was funded by a Japan Society for the Promotion of Science (JSPS) Grants-in-Aid for Early Career Scientists (KAKENHI WAKATE) grant (no. 21K18005) and JSPS Grants-in-Aid for Scientific Research B (KAKENHI Kiban B) grant (no. 24K02972), as well as a Japan Science and Technology Agency (JST) ACT-X grant (no. JPMJAX22A3). We thank Suzuka Yoshida and the members of the Aspirational Computing lab for pilot testing and research assistance.
\end{acks}

\bibliographystyle{ACM-Reference-Format}
\begin{CJK}{UTF8}{ipxm}
\bibliography{REFS}
\end{CJK}




\end{document}